\documentclass[prl,aps,twocolumn,showpacs,reprint,amsmath,amssymb,floatfix]{revtex4-2}
\usepackage{amsmath}
\usepackage{graphicx}
\usepackage{dcolumn}
\usepackage{bm}
\usepackage{float}
\usepackage{xcolor}
\usepackage{lineno}
\usepackage{subfigure}
\usepackage{ulem}
\usepackage{multirow}
\usepackage[colorlinks=true,linkcolor=blue]{hyperref}
\usepackage{silence}
\UseRawInputEncoding
\begin{document}

\preprint{APS/123-QED}

\title{Persistence of Layer-Tolerant Defect Levels in ReS$_2$}

\author{Nikhilesh Maity}
\altaffiliation{These authors contributed equally to this work.}
\author{Shibu Meher}
\altaffiliation{These authors contributed equally to this work.}
\author{Manoj Dey}
\author{Abhishek Kumar Singh}
\email {abhishek@iisc.ac.in}
\affiliation{Materials Research Centre, Indian Institute of Science, Bangalore 560012, India}

\renewcommand{\thefootnote}{\fnsymbol{footnote}}

\date{\today}

\begin{abstract}
Defects in two-dimensional (2D) semiconductors play a decisive role in determining their electronic, optical, catalytic and quantum properties. Understanding how defect energy levels respond to variations in layer thickness is essential for achieving reproducible and scalable device performance. We report the persistence of layer-tolerant defect levels in rhenium disulfide (ReS$_2$), where both donor- and acceptor-type charge transition levels remain nearly unchanged from monolayer to bulk in both AA and AB stacking. The associated two-level quantum system also retains its character across thicknesses, enabling ReS$_2$ to serve as a platform for layer-tolerant single-photon emitters. The invariance arises from the interplay between electronic energy minimization and structural relaxation, which together counteract quantum confinement and reduced dielectric screening. Additionally, the intrinsically weak interlayer coupling in ReS$_2$ plays a crucial role. Our findings uncover the microscopic origin of this unique behavior, distinguishing ReS$_2$ from other transition-metal dichalcogenides and highlighting its potential for thickness-independent optoelectronic and quantum photonic applications.

\end{abstract}

\maketitle

\section{Introduction}
In two-dimensional (2D) semiconductors, defects play a pivotal role in tuning their properties for next-generation electronics, catalysis, quantum technologies, and energy applications.~\cite{Ko2024}. They can act as carrier donors~\cite{Singh2019}, charge traps~\cite{Jiang2018}, scattering~\cite{Chen2024}, recombination centers~\cite{Zhang2019}, or even quantum emitters~\cite{Turunen2022,Dey2025}, making them key to applications ranging from logic transistors and light-emitting devices to single-photon sources. Because of their atomic thickness, 2D materials are highly sensitive to their surrounding environment and structural stacking order, and the energetics of their defect states can be strongly modulated by layer thickness. This dependence presents both challenges and opportunities for controlling device performance.

2D transition metal dichalcogenides (TMDs) have emerged as promising platforms for next-generation electronic and optoelectronic applications~\cite{Arora2021,Mandal2023,Mishra2024,Mak2010,Wang2012,Radisavljevic2011,Chhowalla2013}. These materials exhibit a characteristic transition from a direct band gap in monolayers to an indirect band gap in bulk form~\cite{Bhattacharyya2012,Tongay2012}, which leads to a pronounced reduction in light emission and absorption efficiency~\cite{Splendiani2010}. For MoS$_2$ and WS$_2$, a similar monolayer-to-bulk transition results in reduced carrier mobility~\cite{Radisavljevic2011}. These properties evolve smoothly with the number of layers, as evidenced by the band-gap narrowing and changes in band dispersion reported for black phosphorus~\cite{Qiao2014} and MoS$_2$~\cite{Zhao2018}. Layer-dependent electronic behavior has also been reported in other classes of van der Waals materials~\cite{Kayyalha2016,Choi2017,Jiao2019}. In many TMDs, the relative position of defect levels with respect to the band edges varies strongly with thickness. For example, in MoS$_2$, shallow donor levels associated with hydrogen impurities become deep states as the number of layers increases, thereby suppressing $n$-type conductivity~\cite{Singh2022}. In encapsulated black phosphorus, increasing the number of surrounding boron nitride layers enhances dielectric screening and reduces defect ionization energies~\cite{Zhu2021}. Furthermore, environmental effects and defect tunability can be controlled through the interplay between thickness and defect location~\cite{Wang2020}. This shift originate from a combination of quantum confinement, changes in dielectric screening, and interlayer coupling.

Such sensitivity complicates reproducibility: devices fabricated on monolayers, which possess distinct properties, may behave differently from those based on few-layer or bulk counterparts. Because properties vary substantially with the number of layers, the desirable characteristics of multilayer ($>$1L) and bulk materials are often not preserved, limiting their applicability in layer-tolerant device architectures. In van der Waals (vdW) layered materials, including TMDs, individual layers are stacked via weak vdW forces, enabling exfoliation down to few layers or monolayers using various experimental methods~\cite{Tongay2014,Manzeli2017,Meng2019,Sun2017,Collins1992,PereaLpez2014,Arora2021,Kim2022}. However, producing uniform monolayers over large areas remains challenging~\cite{Yu2013,Choi2017}. Controlling layer number during exfoliation is difficult, which restricts device performance and hinders scalable industrial applications~\cite{Jiao2019,Rathi2015,Yu2014,Li2017,Hong2016}. Consequently, identifying 2D materials with layer-tolerant properties is essential for advancing next-generation nanoscale devices.

Among TMDs, rhenium disulfide (ReS$_2$) exhibits several intriguing properties. It crystallizes in a triclinic structure with a distorted octahedral coordination around the Re atoms~\cite{Tongay2014natcomm}. This reduced lattice symmetry leads to pronounced linear polarization anisotropy in its optical response~\cite{Chenet2015}. Notably, the band gap remains nearly invariant with thickness, and bulk ReS$_2$ behaves as electronically and vibrationally decoupled layers~\cite{Tongay2014natcomm,Jariwala2016}. This characteristic, often termed as ``layer-decoupling'', is attributed to a Peierls-like distortion driven by Re–Re intermetallic bonding~\cite{Tongay2014natcomm}, which suppresses interlayer coupling and preserves low-dimensional properties even in the bulk phase~\cite{Feng2015}. Moreover, the direct band gap nature persists across stacking orders (AA and AB) regardless of layer number~\cite{Zhou2020}. In AB stacking, the difference between direct and indirect band gaps in multilayer ReS$_2$ (2L and 4L) is negligible (less than 0.01 eV), while in AA stacking, the band gap remains direct throughout the monolayer-to-bulk evolution. Optical absorption and Raman spectra show negligible changes under hydrostatic pressure modulation of interlayer distance, indicating very weak interlayer interactions~\cite{Tongay2014natcomm}. Due to this significantly low interlayer coupling strength (ICS), ReS$_2$ exhibits layer-intolerant electronic, optical, and vibrational properties. However, the defect thermodynamics of ReS$_2$ remains largely unexplored and is expected to differ markedly from other TMDs.

Using density functional theory (DFT), we investigate the origin of the layer-tolerant behavior of defect levels in ReS$_2$. We find that Re vacancies (V$_{Re}$) and Re$_{S2}$ antisite defects behave as amphoteric defects, in contrast to the electrically inactive sulfur vacancies V$_{S1}$ and V$_{S2}$. Importantly, the defect charge transition levels remain nearly constant as the system dimensionality changes from monolayer to bulk for both AA and AB stacking. This layer tolerance of defect levels is rationalized using a constrained charge transfer scheme within a jellium framework. We propose that, in addition to dielectric screening effects (SE) and quantum confinement effects (QCE), the weak interlayer coupling strength (ICS) plays a crucial role in ReS$_2$. The exceptionally weak interlayer coupling is identified as the key origin of defect level tolerance, making ReS$_2$ a promising candidate for future layer-tolerant optoelectronic applications.

\section{Methodology}
First-principles calculations were performed using density functional theory (DFT) as implemented in the \textit{Vienna Ab-initio Simulation Package} (VASP)~\cite{Kresse1996,Kresse1996comp}. The projector augmented wave (PAW) method~\cite{Blochl1994,Kresse1999} was employed to treat valence and core electrons. Exchange-correlation effects were described using the Perdew-Burke-Ernzerhof (PBE) generalized gradient approximation (GGA)~\cite{Perdew1996}. Kohn-Sham orbitals were expanded in a plane-wave basis with a 500 eV energy cutoff. To avoid spurious interactions between periodic images along the crystallographic $c$ axis, a vacuum spacing of 20 \textup{\AA} was applied for multilayer ReS$_2$. Van der Waals interactions were included via the optB86b-vdW functional~\cite{Klime2009,Klime2011}. The Brillouin zone was sampled with a $16\times16\times1$ Monkhorst-Pack $k$-point mesh for 2D multilayer primitive cells and $16\times16\times8$ for bulk ReS$_2$ primitive cells~\cite{Monkhorst1976}. All structures were fully relaxed using the conjugate-gradient algorithm until forces and total energies converged below 0.005 eV/$\AA$ and $10^{-6}$ eV, respectively. Point defects were modeled within the supercell approach~\cite{Freysoldt2014,VandeWalle2004}, using a $3\times3\times1$ supercell (108 atoms) for the monolayer and $N \times 108$ atoms for multilayers with $N$ layers. Defect calculations employed $2\times2\times1$ and $2\times2\times2$ $k$-point meshes for 2D multilayer and bulk ReS$_2$, respectively.

\begin{figure*}[t]
\begin{center}
\includegraphics[width=0.9\textwidth]{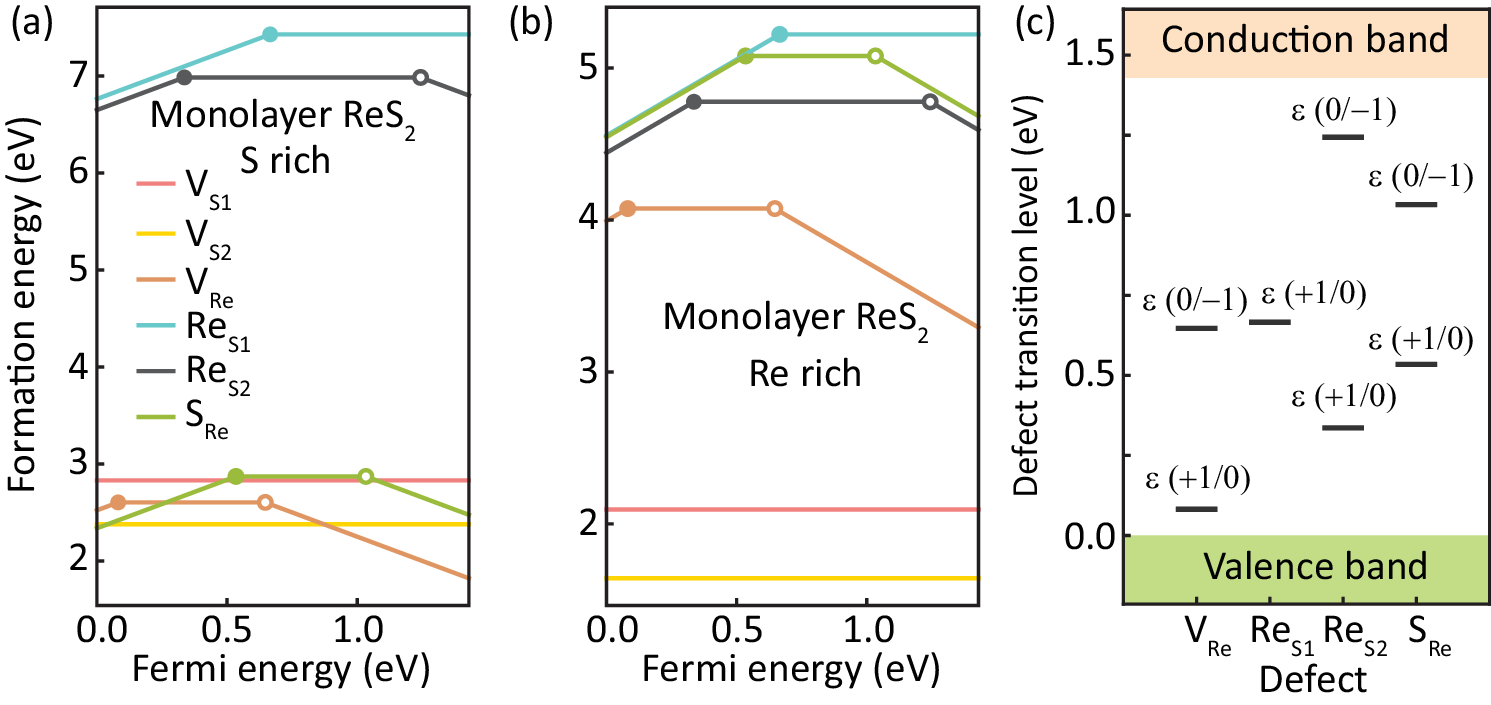}
\end{center}
\caption{Formation energy of intrinsic point defects in monolayer ReS$_2$ as a function of Fermi level under (a) S-rich and (b) Re-rich conditions. The acceptor and donor levels are represented as solid and hollow circles, respectively. (c) Donor [$\epsilon$(+1/0)] and acceptor [$\epsilon$(0/-1)] charge transition levels for point defects in monolayer ReS$_2$. The valence band and conduction band energy regions are plotted in green and orange, respectively.}
\label{fig:fig1}
\end{figure*}

In order to assess the thermodynamics of defects, we calculate their formation energies, which quantify the energetic cost to create a defect in the system. For a defect $X$ in charge state $q$, the formation energy is given by~\cite{Freysoldt2014,Dey2021,Dey2023}:
\begin{equation}
E^f[X^q] = E^{Defect}_{tot} - E^{Pristine}_{tot} - \sum_i n_i \mu_i + qE_F + \Delta^q
\label{eq:1}
\end{equation}

where $E^{Defect}_{tot}$ and $E^{Pristine}_{tot}$ are the total energy of the supercell with and without the defect $X$, respectively. $n_i$ represents the number of atoms of the $i^{th}$ species that are added ($n_i > 0$) or removed ($n_i < 0$) from the system and $\mu_i$ is the chemical potential of $i^{th}$ species. $\mu_i$ represents the energy of the reservoir referenced to the respective standard phases. For Re and S, the references are the energy of pure  Re ($\mu_{Re}^{bulk}$) and S ($\mu_{S}^{bulk}$), respectively. In equilibrium, the chemical potentials of Re and S are related to the formation enthalpy of ReS$_2$ ($\Delta H_{ReS_2}$).

\begin{equation}
\Delta H_{ReS_2} = \mu_{ReS_2}-\mu_{Re}^{bulk} - \mu_S^{bulk}
\label{eq:2}
\end{equation}

also 

\begin{equation}
\mu_{ReS_2} = \mu_{Re} + \mu_S
\label{eq:3}
\end{equation}

In order to avoid the formation of the bulk phase of Re and S, the value of chemical potential is restricted ($\mu_{i} \leq \mu_{i}^{bulk}; i = Re, S$)~\cite{Freysoldt2014}. The Fermi level ($E_F$) represents the chemical potential of electrons and is aligned to the valence-band maximum (VBM). $\Delta^q$ is the finite-size charge correction for the artificial electrostatic interaction between periodic supercells~\cite{Freysoldt2009,Freysoldt2010}. The charge correction for the defect is performed using the FNV scheme~\cite{Freysoldt2009,Freysoldt2018}. 

In order to estimate the deep or shallow nature of the defect, we have calculated the defect transition level (DTL), which is defined as the position of the Fermi level at which a defect changes its charge state from $q$ to $q^\prime$~\cite{Freysoldt2014}. The DTL of defect $X$ is 

\begin{equation}
\varepsilon (q/q^\prime) = \frac{E^f[X^q ; E_F = 0]- E^f[X^{q^\prime} ; E_F =0]}{q^\prime - q}
\label{eq:5}
\end{equation}

The position of $\varepsilon (q/q^\prime)$ from the band edges, i.e., VBM/CBM determines the deep or shallow nature of any defect (acceptor/donor defect).

\section{Results}
\subsection{Unique Electronic Properties of ReS$_2$}
Besides its novel anisotropic properties, ReS$_2$ exhibits a significantly low but finite interlayer coupling strength~\cite{Zhou2020,Zhou2021,Upadhyay2022}, distinguishing it from other TMDs in terms of electronic, optical, and vibrational behavior. The effect of dimensionality on the band gap is minimal: for AA stacking, the band gap increases from 1.22 to 1.33 eV, and for AB stacking, from 1.24 to 1.36 eV, when reducing dimensionality from bulk (3D) to monolayer (2D) as shown in Fig. S1 of Supplemental Material (SM)~\cite{SM} for details, see also Refs.~\cite{Dileep2016,Tongay2014natcomm,Jariwala2016} therein. To assess whether the observed layer-independence is sensitive to the choice of electronic-structure method, we systematically examined the band-gap evolution of ReS$_2$ as a function of layer number using PBE, HSE06, G$_0$W$_0$, and G$_0$W$_0$+BSE approaches. The calculated band gaps for monolayer, bilayer (AA and AB stackings), and bulk (AA and AB stackings) ReS$_2$ are summarized in Table S1 of SM~\cite{SM}. The PBE calculated band gap values and their nature agree well with previous experimental reports~\cite{Zhou2020}. Dimensionality reduction induces two main effects: enhanced quantum confinement and reduced dielectric screening. Upon lowering dimensionality, the valence band maximum (VBM) shifts downward and the conduction band minimum (CBM) shifts upward for both AA and AB stacking (see Fig. S1 in SM~\cite{SM}), resulting in an increased band gap. However, the magnitude of these shifts and the band gap variation are notably small, indicating weak quantum confinement in ReS$_2$. Conversely, the dielectric constant and thus screening decreases rapidly from bulk to monolayer in both AA [Fig.~\ref{fig:fig4}(a)] and AB stacking [Fig. S7(a) in SM~\cite{SM}], a trend more pronounced than in other TMDs~\cite{Singh2022}. The contrasting behavior of quantum confinement and dielectric screening across dimensions is expected to significantly influence the defect thermodynamics in ReS$_2$.

\subsubsection{Point Defects in Monolayer ReS$_2$}
The lower lattice symmetry of ReS$_2$ results in two crystallographically inequivalent sulfur sites, $S1$ and $S2$, which are considered for vacancy ($V_{S1}$, $V_{S2}$) and antisite ($Re_{S1}$, $Re_{S2}$) defects. In addition, rhenium vacancies ($V_{Re}$) and antisites where sulfur substitutes for rhenium ($S_{Re}$) are also studied. The relaxed atomic geometries of these defects in the neutral charge state are shown in Fig. S2(a)–(f) in SM~\cite{SM}. Fig.~\ref{fig:fig1}(a) and Fig.~\ref{fig:fig1}(b) present the formation energies of native point defects as functions of the Fermi level ($E_F$), under sulfur-rich and rhenium-rich conditions, respectively. Among the considered defects, vacancies ($V_{S1}$, $V_{S2}$, and $V_{Re}$) exhibit lower formation energies than antisites under both growth environments, indicating their preferential formation and higher equilibrium concentrations. Under sulfur-rich conditions, $V_{Re}$ is the most stable defect in the $-1$ charge state over a wide range of $E_F$ near the CBM. Sulfur vacancies preferentially form in rhenium-rich environments, while rhenium vacancies are favored in sulfur-rich growth. The S$_{Re}$ antisite exhibits significantly lower formation energy in sulfur-rich conditions compared to rhenium-rich conditions, indicating its easier formation under the former. In contrast, antisites $Re_{S1}$ and $Re_{S2}$ have high formation energies in both growth conditions, attributed to the large size mismatch that disfavors incorporation of Re atoms at sulfur sites.

\begin{figure*}[t]
\begin{center}
\includegraphics[width=0.8\textwidth]{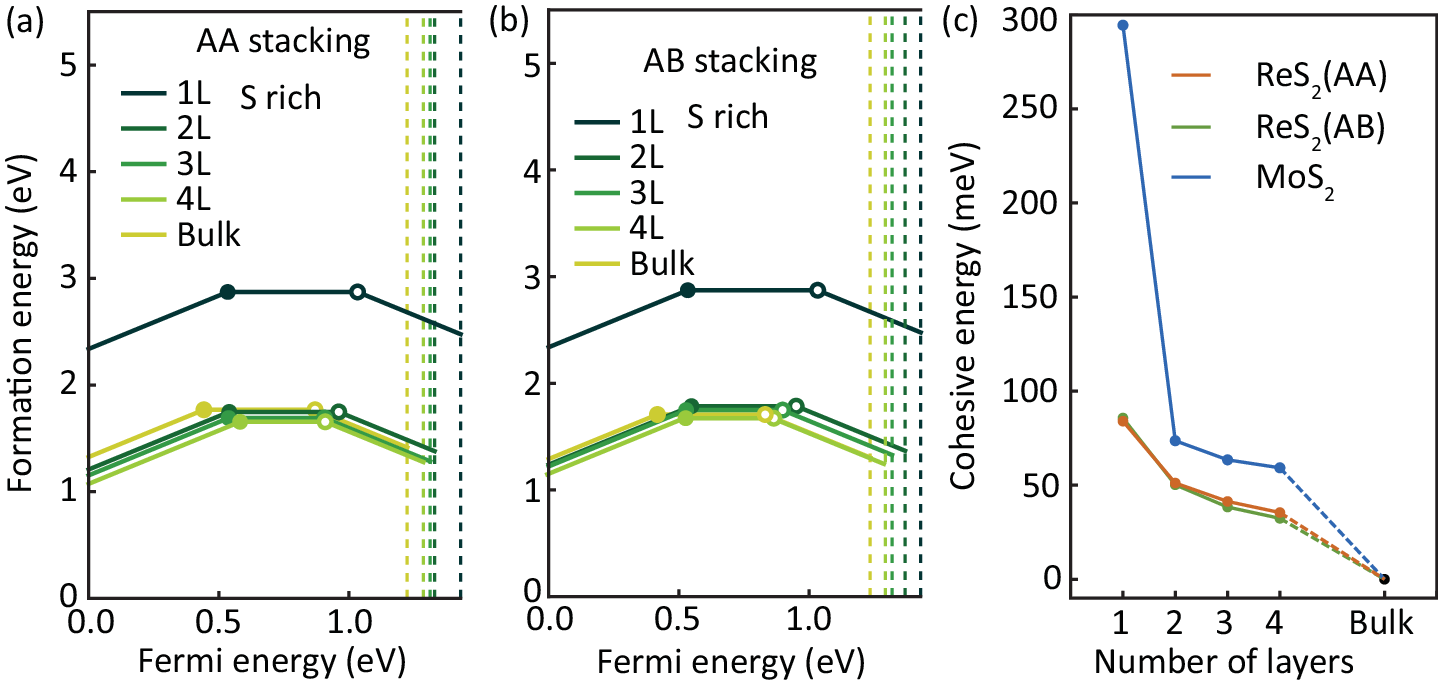}
\end{center}
\caption{Formation energy as a function of Fermi level of S$_{Re}$ defect in 1L to Bulk ReS$_2$ for (a) AA stacking and (b) AB stacking under S-rich conditions. (c) The cohesive energy of layered ReS$_2$ for AA, AB stacking and layered MoS$_2$ as a function of dimensionality.}
\label{fig:fig2}
\end{figure*}

\subsubsection{Defect Transition Levels of Point Defects}
To further elucidate the stability of defect charge states, the corresponding charge transition levels are presented in Fig.~ \ref{fig:fig1}(c). Sulfur vacancies ($V_{S1}$, $V_{S2}$) remain stable in the neutral charge state over the entire Fermi level range, as shown in Fig.~\ref{fig:fig1}(a) and \ref{fig:fig1}(b); consequently, no charge transition levels appear within the band gap for these defects. The $Re_{S1}$ antisite acts as a deep hole trap, with its donor transition level ($0/+1$) located 0.67 eV above the VBM. $V_{Re}$, $Re_{S1}$, and $Re_{S2}$ defects exhibit stable charge states of $+1$, 0, and $-1$. Notably, their donor and acceptor levels are deep, isolated from the band edges and thus robust against external perturbations. For $V_{Re}$, the ($+1/0$) and ($0/-1$) transition levels lie 1.35 eV and 0.78 eV below the CBM, respectively. This results in donor behavior from the VBM up to $\varepsilon(+1/0)$, neutral charge between $\varepsilon(+1/0)$ and $\varepsilon(0/-1)$, and acceptor behavior from $\varepsilon(0/-1)$ to the CBM. Similarly, for $Re_{S2}$, the ($+1/0$) and ($0/-1$) levels are located 1.09 eV and 0.18 eV below the CBM. The $S_{Re}$ antisite exhibits transition levels ($0/-1$) at 0.90 eV below the CBM and ($+1/0$) at 1.03 eV above the VBM. This placement of defect levels reduces the influence of the host band edges on its defect states, preventing the ionization of neutral charge state.

\subsection{Dimensionality Effect on the Defect Thermodynamics}
To elucidate the effect of dimensionality on defect thermodynamics, we focus on the prominent $S_{Re}$ antisite defect. A prerequisite is to determine the actual stacking order of layered ReS$_2$. Due to the extremely weak interlayer coupling, the literature reports conflicting results regarding ReS$_2$ stacking~\cite{Lin2015,Aslan2015,Qiao2016}. Recently, two energetically favorable stacking configurations AA and AB have been identified based on translational displacements, showing good agreement with experimental observations~\cite{Zhou2020}. Both AA and AB stacking sequences are considered here to assess their influence on defect behavior as a function of dimensionality. For multilayer systems, the position of the defective layer critically affects defect thermodynamics due to varying dielectric screening environments from adjacent layers~\cite{Wang2020}. To maximize screening effects, the defect is always introduced in an inner layer for systems with $N > 2$ (see Fig. S3). To capture the dimensionality dependence, we calculate the formation energy of $S_{Re}$ for monolayer (1L, $N=1$, ideal 2D), bilayer (2L, $N=2$, quasi-2D), trilayer (3L, $N=3$, quasi-2D), four-layer (4L, $N=4$, quasi-2D), and bulk ($N \to \infty$, 3D) ReS$_2$. The evolution of $S_{Re}$ formation energy from 2D to bulk is presented in Fig.~\ref{fig:fig2}(a) and \ref{fig:fig2}(b).

The formation energies of $S_{Re}$ for AA stacking under sulfur- and rhenium-rich conditions are shown in Fig.~\ref{fig:fig2}(a) and Fig. S4(a) in SM~\cite{SM}, respectively. As expected, the formation energy in multilayer and bulk ReS$_2$ is lower under sulfur-rich conditions than under rhenium-rich conditions, consistent with the monolayer case. Notably, a substantial reduction in formation energy is observed as dimensionality increases from monolayer to bulk. For instance, the formation energy of the $+1$ charge state at VBM decreases from 2.34 eV in the monolayer to 1.32 eV in the bulk for AA stacking under sulfur-rich growth. Similarly, for AB stacking under sulfur-rich conditions, it decreases from 2.34 eV to 1.29 eV [see Fig. \ref{fig:fig2}(b) and S4(b) in SM~\cite{SM}]. Comparable trends are observed under rhenium-rich conditions. This variation in formation energy correlates with the layer-dependent cohesive energy of ReS$_2$, which is quantified by the following expression:

\begin{equation}
E_{cohesive} = \frac{E_{NL}}{12N} - \frac{E_{bulk}}{24}
\label{eq:6}
\end{equation}

where, $E_{NL}$ is the total energy of the N layer ReS$_2$ and $E_{bulk}$ is energy of bulk ReS$_2$. There are total 12$\times$N atoms of the N layer ReS$_2$ and a total of 24 atoms in bulk ReS$_2$. The cohesive energy per atom is calculated by normalizing the total energies $E_{NL}$ and $E_{bulk}$ by factors of 12$\times$N and 24, respectively. Here, the bulk ReS$_2$ energy (AA or AB stacking) is taken as the reference to determine the cohesive energy of an N-layer (multilayer) system with the same stacking order. Thus, the cohesive energy of multilayer ReS$_2$ quantifies the energetic cost of forming that multilayer from its bulk phase. As dimensionality decreases from 3D to 2D, the cohesive energy increases for both AA and AB stacking orders, as shown in Fig.~\ref{fig:fig2}(c). It implies that creating thinner layers requires progressively more energy, reaching a maximum for the monolayer. Consequently, defect formation in lower-dimensional regimes demands higher energy costs. Similar trends have been reported in other layered materials~\cite{Singh2022,Wang2020}. To further illustrate this, we extracted the formation energy of the $S_{Re}$ defect in the $+1$ charge state at the VBM for layers ranging from monolayer to bulk under sulfur- and rhenium-rich conditions, shown in Fig.~S5(a) and S5(b) in SM~\cite{SM}, respectively. The formation energies for the neutral and $-1$ charge states exhibit similar dimensionality dependence at fixed Fermi level positions. In all cases, the formation energy saturates beyond a critical thickness (more than three layers for both AA and AB stacking), defining the bulk limit. Since the defective layer is positioned in the inner region, the dielectric screening increases sharply as additional defect-free layers are added on both sides.

The defect transition levels of the $S_{Re}$ antisite across dimensionalities are shown in Fig.~\ref{fig:fig3}(a) and Fig.~S11(a) for the AA and AB stacking orders, respectively. Both donor ($+1/0$) and acceptor ($0/-1$) transition levels lie within the band gap from monolayer to bulk ReS$_2$. As illustrated in Fig.~\ref{fig:fig3}(a), these defect transition levels become shallower with increasing dimensionality. This trend also observed in black phosphorus, MoS$_2$, WS$_2$, and h-BN~\cite{Wang2020,Singh2022,Zhu2021}. However, in ReS$_2$, the changes in defect transition levels are nearly negligible, attributed to its exceptionally weak quantum confinement effects. This fundamental aspect and its origin will be discussed in detail later. Importantly, the defect transitions remain deep in ReS$_2$ across dimensionalities, whereas in other materials, defects tend to shift from deep in 2D to shallow in the 3D limit. Deep defect levels indicate that specific charge states of the defects are highly stable, requiring substantial energy to change charge states. This energy cost is quantified as the ionization energy, defined by the defect transition energy relative to the band edges. Specifically, the ($+1/0$) transition level measured from the CBM corresponds to the donor ionization energy, while the ($0/-1$) transition level measured from the VBM corresponds to the acceptor ionization energy.

\begin{figure*}[htb]
\begin{center}
\includegraphics[width=1.0\textwidth]{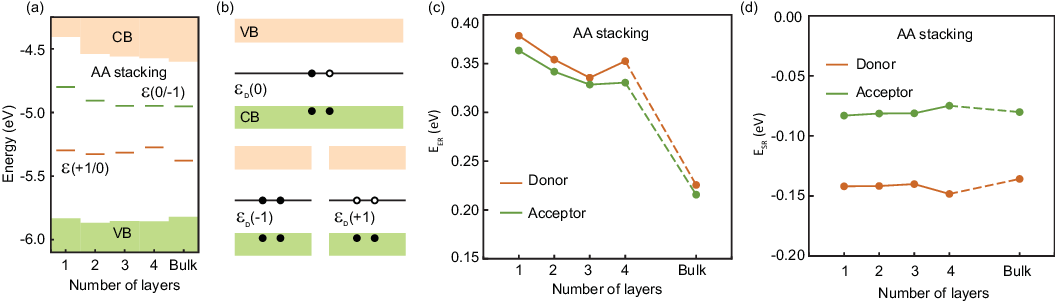}
\end{center}
\caption{(a) Defect transition-energy levels of S$_{Re}$ as function of number of layers for AA stacking. (b) Schematic of electron occupations for a donor and acceptor for the neutral, negative, and positive charge state. (c) Electronic relaxation energy and (d) structural relaxation energy for AA stacking.}
\label{fig:fig3}
\end{figure*}

Figs~S6(a) and S6(b) in SM~\cite{SM} show the variation of the donor and acceptor ionization energies of $S_{Re}$ defects in ReS$_2$ as a function of layer number for AA and AB stacking, respectively. Notably, the stepwise shifts in ionization energy from monolayer to bilayer, bilayer to trilayer, and so forth, do not directly correspond to the band edge shifts. This is because the magnitudes of band edge variations differ from those of the defect transition level variations with layer number. For example, in AA stacking, the CBM and VBM shift by 0.13 eV and 0.03 eV, respectively, from monolayer to bilayer. In AB stacking, these shifts are 0.10 eV (CBM) and 0.04 eV (VBM). The energy difference of ($+1/0$) donor and ($0/-1$) acceptor defect transition levels shift by 0.30 eV and 0.11 eV, respectively, in AA stacking, and by 0.30 eV and 0.13 eV in AB stacking over the same interval. This difference is responsible for the relaxation energy of electrons from $0$ to $+1$ or $0$ to $-1$ charge state, which will be thoroughly examined in the next section. Consequently, the ionization energies exhibit a slight, nonuniform variation from monolayer to bulk, with the largest changes occurring between monolayer and bilayer. Specifically, the donor and acceptor ionization energies change by 0.10 eV and 0.07 eV for AA stacking, and 0.07 eV and 0.08 eV for AB stacking, respectively, between monolayer and bilayer. Beyond this, the ionization energies decrease marginally and saturate toward bulk values, as shown in Figs.~S6(a) and S6(b). Overall, the donor and acceptor ionization energies remain large (corresponding to deep defect levels), indicating that the deep character of $S_{Re}$ defects is essentially independent of dimensionality in ReS$_2$. This stability supports the persistence and robustness of the associated charge states across different thicknesses.

In order to gain insight into the defect properties for specific charge states, we plot the energy level diagrams of various defects in monolayer ReS$_2$ in Fig.~S8 of the SM~\cite{SM}. Each defect is presented in a separate section, with the energy levels corresponding to different charge states organized in subsections. All defect states and band edge positions are aligned with respect to the vacuum level. Upon defect formation, multiple occupied and unoccupied states appear within the band gap. These states arise from the nominal valence electrons of the defect as well as the valence electrons of neighboring atoms~\cite{Wei2002}. Among all considered point defects, unoccupied energy levels appear within the band gap for all $0$, $+1$, and $-1$ charge states of $V_{Re}$. As shown in Fig.~S8 in SM~\cite{SM}, the neutral charge states of $S_{Re}$, $Re_{S1}$, $Re_{S2}$, and $V_{Re}$ each possess odd numbers of electrons, resulting in paramagnetic spin configurations. Furthermore, some of these states introduce occupied and unoccupied levels within the bandgap, forming two-level quantum systems. These two-level quantum systems are optically excitable by laser pulses and thus have potential applications as single-photon emitters (SPEs). Among these, the $Re_{S1}$ defect has a small energy separation of 0.06 eV between its occupied ground and unoccupied excited states, limiting its practical use. In contrast, the $S_{Re}$, $Re_{S2}$, and $V_{Re}$ defects have larger gaps of 0.41, 0.44, and 0.49 eV, respectively. Notably, these two-level quantum systems vanish in the $+1$ and $-1$ charge states of $S_{Re}$, $Re_{S2}$, and $V_{Re}$, indicating that the neutral charge state is essential for their stability. This stability is corroborated by the deep charge transition levels shown in Figs.~\ref{fig:fig1}(a) and (b), which confirm that these defects remain stable and electronically isolated from the band edges in the neutral state.

In other layered materials, these distinctive two-level quantum features are typically sensitive to dimensionality; even a change of one or two layers or a variation in stacking order can drastically alter their properties. Since precise exfoliation to obtain a specific number of layers and stacking configuration is experimentally challenging, there is a critical need for systems having robust, layer- and stacking-tolerant quantum behavior. To investigate the dimensionality effect on the two-level quantum system, we calculated the defect energy levels of $S_{Re}$ in ReS$_2$ for both AA (Fig.~S9) and AB (Fig.~S10) stacking orders. Remarkably, we observe that the energy levels remain largely invariant with dimensionality changes from 2D to 3D. Specifically, the occupied and unoccupied defect states in the neutral charge state show minimal shifts relative to the band edges, and the energy gaps between these levels remain nearly constant across monolayer, bilayer, trilayer, four-layer, and bulk systems. For AA stacking, the gap values are 0.41, 0.39, 0.40, 0.39, and 0.40 eV, respectively, while for AB stacking, they are 0.41, 0.38, 0.39, 0.39, and 0.40 eV. This insensitivity to both dimensionality and stacking order underscores the exceptional layer-tolerant nature of the quantum defect states in ReS$_2$.

\section{Discussion}
From the preceding sections, we identify three key findings regarding the dimensionality effects in layered ReS$_2$. First, the band edge positions exhibit only minor shifts when transitioning from bulk to monolayer. Second, the variation in defect ionization energies with dimensionality is notably small. Although defect transition levels typically evolve from deep to shallow with increasing thickness, the changes in ReS$_2$ are negligible compared to other layered materials~\cite{Zhu2021,Singh2022}. Third, defect energy levels in ReS$_2$ demonstrate pronounced tolerance to dimensionality. The two primary factors that govern property changes from bulk to low dimensional materials are the quantum confinement effect (QCE) and the reduction in dielectric screening effect (SE)~\cite{Zhu2021,Singh2022,Dey2022}. Here, we introduce an additional critical parameter: the interlayer coupling strength (ICS), which plays a significant role in layered ReS$_2$. To connect these factors and explain the dimensionality dependence of charge transition levels, we employ a constrained charge transfer scheme within the Jellium model framework~\cite{Zhu2020}. Within this approach, the defect transition energy level of acceptor (q$<$0) can be expressed with respect to VBM as:

\begin{equation}
\varepsilon (0/q) = E_{NDL} + E_{ER} + E_{SR}
\label{eq:7}
\end{equation}

where, $E_{NDL}$ is the single-electron neutral defect levels. The expression of $E_{NDL}$ is given by,

\begin{equation}
E_{NDL} = \varepsilon^{neu}_D - \varepsilon^{host}_{VBM}
\label{eq:8}
\end{equation}

for a acceptor, where, $\varepsilon^{neu}_D$ is the energy of neutral defect level (Kohn Sham level) and $\varepsilon^{host}_{VBM}$ is the VBM of the host supercell. For acceptor, $\varepsilon^{neu}_D$ is measured from VBM. The transition energy level of donor (q$>$0) can be expressed with respect CBM as,

\begin{equation}
E_g - \varepsilon (0/q) = E_{NDL} + E_{ER} + E_{SR}
\label{eq:7a}
\end{equation}

where $E_g$ is the band gap. The expression of $E_{NDL}$ is given by,

\begin{equation}
E_{NDL} = \varepsilon^{host}_{CBM} - \varepsilon^{neu}_D
\label{eq:9}
\end{equation}

where, $\varepsilon^{neu}_D$ is measured from CBM. $E_{NDL}$ is the contribution for the neutral charge state of any defect, and the other two terms $E_{ER}$ and $E_{SR}$ due to the extra cost of energy in the transition levels for the charged defect. $E_{ER}$ and $E_{SR}$ represent the electron relaxation and structural relaxation contributions due to the extra charge ($+1$ or $-1$) from its neutral state as shown in Fig. \ref{fig:fig3} (b). The expression of $E_{ER}$ is 
 
\begin{equation}
E_{ER} = E^0(X^q) - E(X^0) + q\varepsilon^{neu}_D
\label{eq:10}
\end{equation} 

and the expression of $E_{SR}$ is

\begin{equation}
E_{SR} = E^{relax}(X^q) - E^0(X^q)
\label{eq:11}
\end{equation} 
where, $E^{relax}(X^q)$ is the relaxed energy of defect ($X$) supercell in charged state $q$. $E^0(X^q)$ is the total energy of the $q$ charged defect when its atomic positions are the same as those in the fully relaxed defect supercell of $0$ charge state. $E(X^0)$ is the energy of fully relaxed $0$ charge state of defect $q$. Among these three quantities, $E_{NDL}$ has the most contribution to the defect transition level~\cite{Zhu2020}. $E_{NDL}$ is largely affected by QCE whereas $E_{ER}$ is strongly related to SE.

\begin{figure}[htb]
\begin{center}
\includegraphics[width=0.5\textwidth]{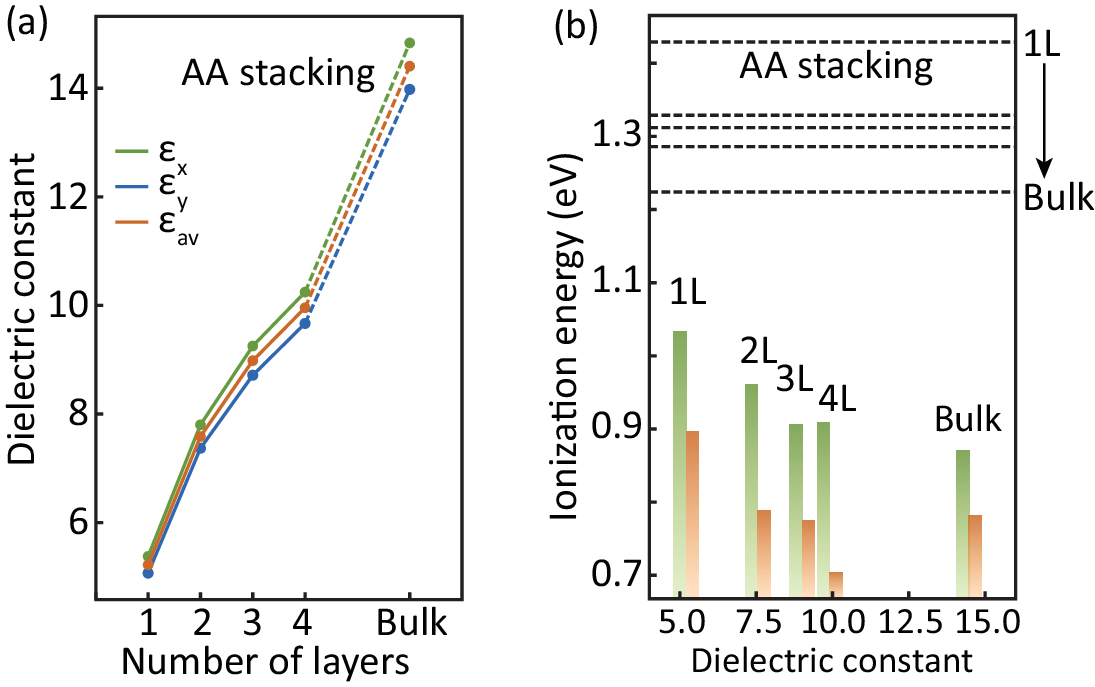}
\end{center}
\caption{(a) Static dielectric constant along x, y directions and average as a function of number of layers of AA stacking order. (b) Ionization energy of S$_{Re}$ as a function of average static dielectric constant for AA stacking order. The green and orange bars represent the ionization energy of acceptors and donors, respectively.}
\label{fig:fig4}
\end{figure}

\subsubsection{Interlayer Coupling Strength (ICS)}
Interlayer coupling strength, alongside quantum confinement effect and screening effect, plays a crucial role in determining the properties of low-dimensional layered materials. Variations in band edges and band gaps are directly influenced by the ICS in layered systems. As shown in Fig.~S1(a) and (b), ReS$_2$ exhibits minimal dimensionality-induced changes in band gap and band edge positions compared to other layered materials such as MoS$_2$ and black phosphorus~\cite{Wang2020,Zhu2021}. To elucidate this behavior, we investigated the cohesive energy from monolayer to bulk ReS$_2$ for both AA and AB stacking orders, presented in Fig.~\ref{fig:fig2}(c). The most significant change in cohesive energy occurs during the monolayer-to-bilayer transition, decreasing from 84.15 to 51.15 meV for AA stacking and from 84.02 to 50.21 meV for AB stacking. Overall, the variation in cohesive energy from monolayer to bulk in ReS$_2$ is substantially smaller than in other TMDs such as MoS$_2$~\cite{Singh2022}, as illustrated in Fig.~\ref{fig:fig2}(c). This indicates that the ICS in ReS$_2$ is significantly weaker than in MoS$_2$, resulting in minimal interlayer orbital interactions. Consequently, band splitting is also limited in ReS$_2$, leading to negligible shifts in band edges and band gaps with changing dimensionality.

\subsubsection{Quantum Confinement Effect (QCE) and Screening Effect (SE)}
In order to explain the minimal change in defect transition levels with dimensionality [shown in Fig.~\ref{fig:fig3}(a)], we analyze the contributions in Eq.~\ref{eq:7} and Eq. ~\ref{eq:7a}. The term $E_{NDL}$ is influenced by the quantum confinement effect. From Fig.~S9 and S10 of SM~\cite{SM}, the defect levels relative to the band edges remain essentially unchanged, indicating a negligible QCE in ReS$_2$. The other two terms, $E_{ER}$ and $E_{SR}$, contribute small but finite to the defect transition levels~\cite{Zhu2020}. Their variation from monolayer to bulk is shown in Fig.~\ref{fig:fig3}(c)-(d) and S11(b)-(c) in SM~\cite{SM}. Here, $E_{SR}$ represents the structural relaxation energy associated with changing the defect charge state, and for $S_{Re}$ it remains nearly constant at approximately -0.14 eV for the donor and 0.09 eV for the acceptor across all thicknesses and stacking orders. In contrast, $E_{ER}$ denotes the electronic relaxation energy cost accompanying the charge state transition, which is directly related to the screening properties of the system. In the 3D bulk, the Coulomb interaction of charged defects is effectively screened by surrounding atoms, whereas in 2D, screening is limited to in-plane neighbors within a few layers. Reduced screening in the 2D regime leads to a higher electronic relaxation energy cost compared to bulk.

The variation of the electronic relaxation energy, $E_{ER}$, with dimensionality is shown in Fig.~\ref{fig:fig3}(c) for AA stacking and Fig. S11(b) for AB stacking, for both donor and acceptor defects. In both stacking orders, $E_{ER}$ increases as the number of layers decreases from bulk to monolayer. Specifically, for the donor state, $E_{ER}$ rises from 0.23 eV in bulk AA stacking and 0.22 eV in bulk AB stacking to 0.38 eV in the monolayer. For the acceptor state, $E_{ER}$ increases from 0.20 eV (bulk AA) and 0.21 eV (bulk AB) to 0.36 eV in the monolayer. The positive $E_{ER}$ increases the defect transition level $\varepsilon(0/q)$, while the negative structural relaxation energy $E_{SR}$ (approximately -0.14 eV) offsets it by a nearly constant amount across dimensionalities. Overall, this leads to a decrease in $\varepsilon(0/q)$, i.e., a transition from deep to shallow defect transition levels from monolayer to bulk, as shown in Fig.~\ref{fig:fig3}(a) and Fig.~S11(a). These results suggest that the reduction in defect ionization energy from monolayer to bulk ReS$_2$ is primarily driven by increased screening effects. To further support this conclusion, we plot the ionization energy of donor and acceptor levels of $S_{Re}$ against the dielectric constant in Fig.~\ref{fig:fig4}(b) for AA stacking and Fig.~S7(b) for AB stacking. This correlation confirms the decrease of defect transition levels with enhanced dielectric screening from monolayer to bulk in both stacking configurations.

Overall, the nearly thickness-independent defect ionization energies observed here indicate that defect states in ReS$_2$ are weakly hybridized with interlayer electronic states. To assess the generality of this behavior, we compare interlayer coupling strengths in structurally related materials. ReSe$_2$ exhibits a slightly larger interlayer coupling than ReS$_2$, yet both values remain substantially smaller than those of conventional transition-metal dichalcogenides (see Fig. S12 in SM~\cite{SM}). Consistent with this weak interlayer interaction, the band gap of ReSe$_2$ shows only a modest dependence on layer number, with experimentally reported reductions of $\sim$0.1-0.13 eV from monolayer to bulk~\cite{Arora2017,Jariwala2016}, closely resembling the behavior of ReS$_2$. While these similarities suggest that defect ionization energies in ReSe$_2$ may also exhibit weak thickness dependence, a detailed defect-level analysis is required to confirm this expectation. In contrast, materials such as MoS$_2$ and WS$_2$ exhibit significantly stronger interlayer coupling and much larger band-gap variations with thickness, which have been shown to lead to pronounced layer-dependent defect ionization energies~\cite{Singh2022,Zhu2021}.

\section{Conclusion}
In conclusion, we have demonstrated the layer-tolerant behavior of defect levels in ReS$_2$. The charge transition levels exhibit negligible shifts from deep to shallow states as the dimensionality changes from monolayer to bulk. Notably, the energy gap between the two-level quantum system remains nearly constant at approximately 0.40 eV and 0.39 eV for AA and AB stacking, respectively. This persistent behavior is explained through a constrained charge transfer scheme within a jellium framework, highlighting the interplay among electronic relaxation energy, and structural relaxation. We attribute the observed tolerance primarily to the exceptionally weak interlayer coupling strength in ReS$_2$, establishing it as a unique platform for robust layer-independent defect properties.

\section{Acknowledgements}
The authors express thanks to the Materials Research Center (MRC) and Supercomputer Education and Research Centre (SERC), as well as the Solid State and Structural Chemistry Unit (SSCU) of the Indian Institute of Science, Bangalore, for granting access to the necessary computing resources. The authors thank the support from the DST-Nanomission programme of the Department of Science and Technology, Government of India (DST/NM/TUE/QM-1/2019). Shibu Meher acknowledges the PMRF fellowship (ID: 0201908). The authors also acknowledge the support from The Institute of Eminence (IoE) scheme of The Ministry of Human Resource Development, Government of India.

%

\clearpage
\widetext
\begin{center}
\textbf{\large Supplemental Material: Persistence of Layer Tolerant Defect Levels in ReS$_2$}
\end{center}

\setcounter{equation}{0}
\setcounter{figure}{0}
\setcounter{table}{0}
\setcounter{page}{1}
\makeatletter
\renewcommand{\theequation}{S\arabic{equation}}
\renewcommand{\thefigure}{S\arabic{figure}}
\renewcommand{\bibnumfmt}[1]{[S#1]}
\renewcommand{\citenumfont}[1]{S#1}


\begin{figure*}[ht!]
\begin{center}
\includegraphics[width=0.5\textwidth]{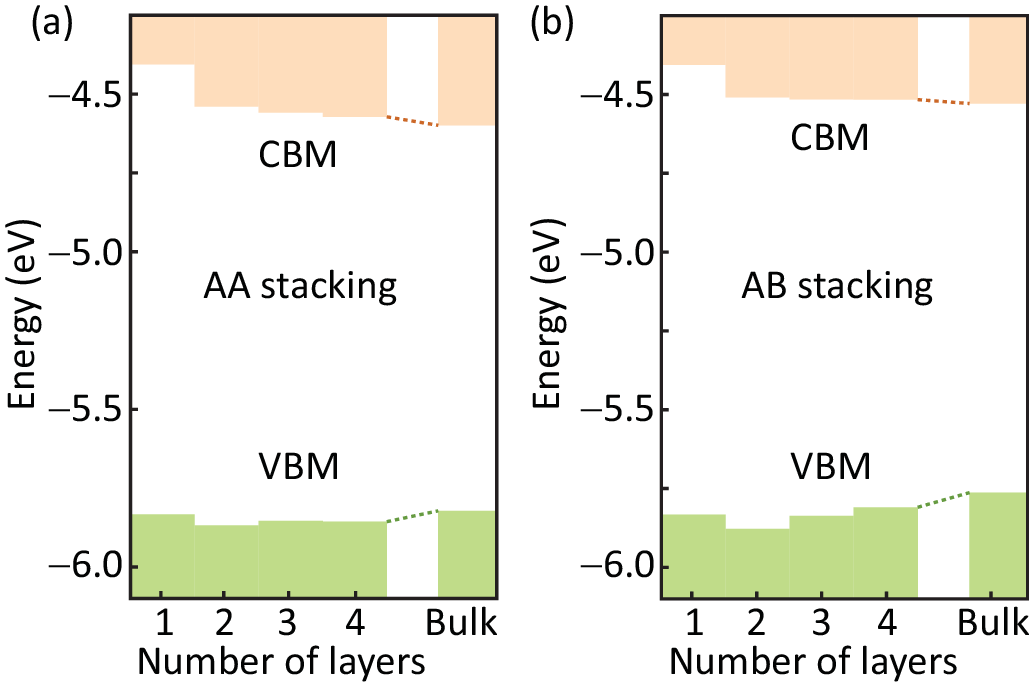}
\end{center}
\caption{Band edge positions with respect to vacuum level of ReS$_2$ for (a) AA stacking and (b) AB stacking configurations, illustrating the evolution from monolayer to bulk.}
\label{fig:fig1}
\end{figure*}

\begin{table*}[th]
\caption{Calculated and experimental band gaps (in eV) of ReS$_2$.}
\centering
\begin{tabular}{|lccccc|}
\hline
System & PBE & HSE06 & G$_0$W$_0$ & G$_0$W$_0$+BSE & Experiment \\
\hline
1L        & 1.42 & 2.06 & 2.15 & 1.70 & 1.51--1.60 ~\cite{Dileep2016,Tongay2014natcomm,Jariwala2016} \\
\hline
2L-AA     & 1.33 & 1.95 & 1.98 & 1.68 & \multirow{2}{*}{1.46--1.51 ~\cite{Dileep2016,Tongay2014natcomm,Jariwala2016}} \\
2L-AB     & 1.36 & 1.99 & 2.01 & 1.70 & \\
\hline
Bulk-AA   & 1.22 & 1.83 & 1.87 & 1.56 & \multirow{2}{*}{1.42--1.55 ~\cite{Dileep2016,Tongay2014natcomm,Jariwala2016}} \\
Bulk-AB   & 1.24 & 1.84 & 1.89 & 1.57 & \\
\hline
\end{tabular}
\label{tab:bandgaps}
\end{table*}

At the PBE level, the band gap decreases from 1.42 eV in the monolayer to 1.22 eV in bulk AA stacking, corresponding to a total variation of 0.20 eV (0.18 eV for AB stacking). Hybrid-functional (HSE06) and quasiparticle G$_0$W$_0$ calculations yield larger absolute band-gap values compared to experiment; however, the thickness-dependent trend remains consistent across all methods. In particular, the reduction in band gap from monolayer to bulk is comparable to that obtained at the PBE level. Importantly, even when excitonic effects are included within the G$_0$W$_0$+BSE framework, the total band-gap variation remains modest, amounting to 0.14 eV (from 1.70 eV to 1.56 eV) for AA stacking and 0.13 eV (from 1.70 eV to 1.57 eV) for AB stacking. These results demonstrate that, although different electronic-structure methods predict different absolute band-gap values, the magnitude of the band-gap change with increasing thickness is small.
Given the substantially higher computational cost of HSE06 and many-body perturbation approaches, the PBE results provide a reasonable description of both the absolute band-gap values and their thickness dependence in comparison with available experimental data, whereas the higher-level methods slightly overestimate the band gaps.

\begin{figure*}[t]
\begin{center}
\includegraphics[width=0.8\textwidth]{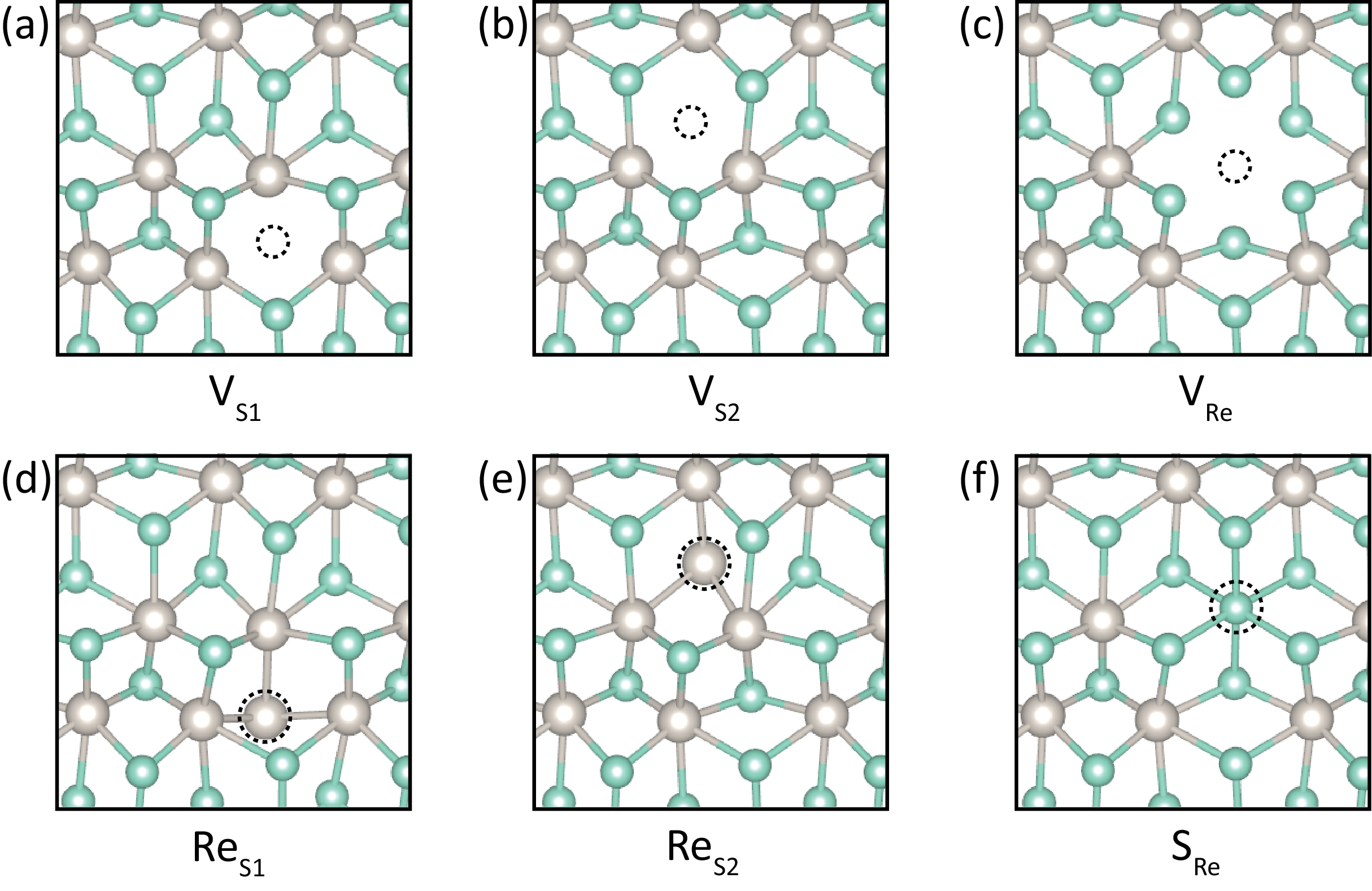}
\end{center}
\caption{Atomic geometry of (a) V$_{S1}$, (b) V$_{S2}$, (c) V$_{Re}$, (d) Re$_{S1}$, (e) Re$_{S2}$ and (f) S$_{Re}$ point defects in the 0 charge states.
The smaller dotted circle in figure (a), (b), and (c) shows the position of the vacancy defect. The bigger dotted circle in figure
(d), (e), and (f) shows the position of the antisite defects.}
\label{fig:fig2}
\end{figure*}

\begin{figure*}[t]
\begin{center}
\includegraphics[width=0.8\textwidth]{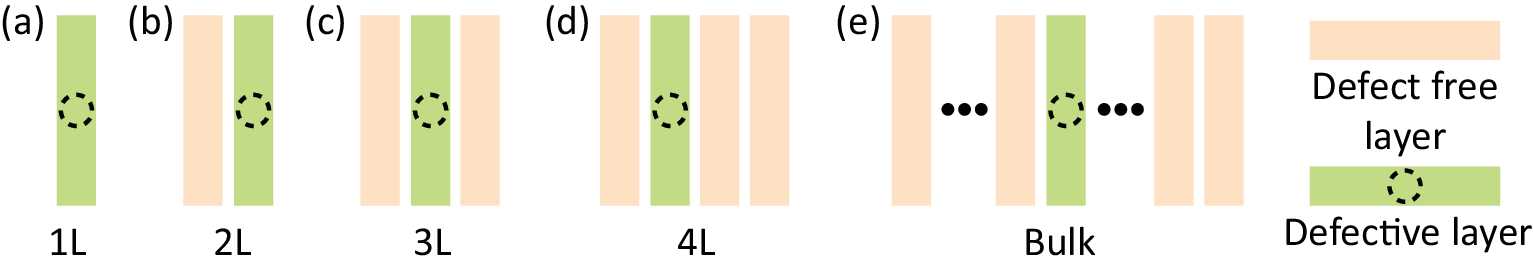}
\end{center}
\caption{The placement of defective layers in multilayer ReS$_2$ of (a) 1L, (b) 2L, (c) 3L, (d) 4L and (bulk). The defective layer and the defect free layer are colored in green and orange, respectively.}
\label{fig:fig3}
\end{figure*}

\begin{figure*}[t]
\begin{center}
\includegraphics[width=0.6\textwidth]{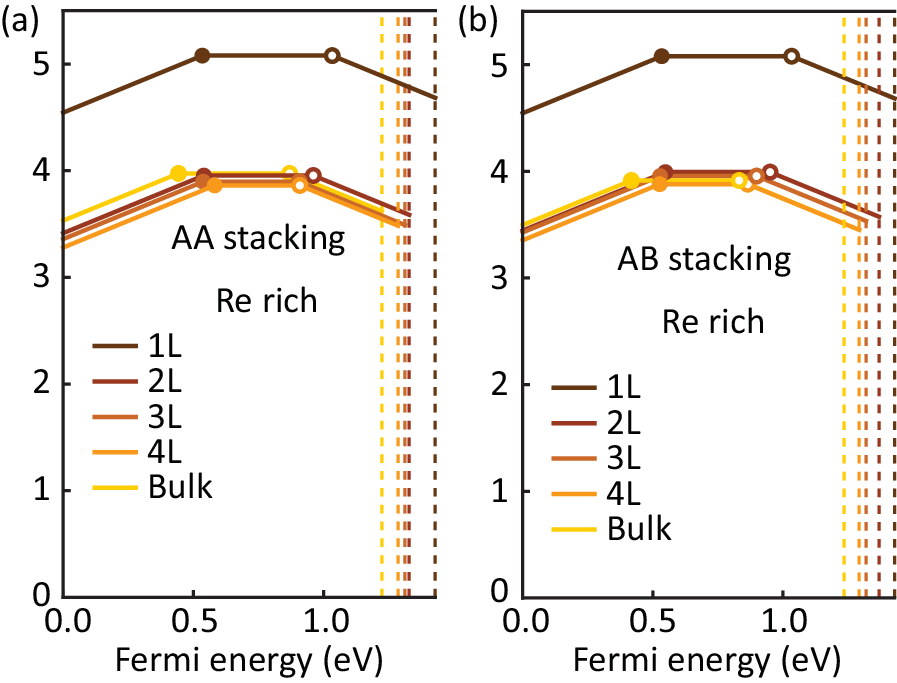}
\end{center}
\caption{Formation energy of S$_{Re}$ in 1L to Bulk ReS$_2$ under Re-rich conditions (a) for AA stacking and (b) for AB stacking as a function of Fermi level.}
\label{fig:fig4}
\end{figure*}

\begin{figure*}[htb]
\begin{center}
\includegraphics[width=0.6\textwidth]{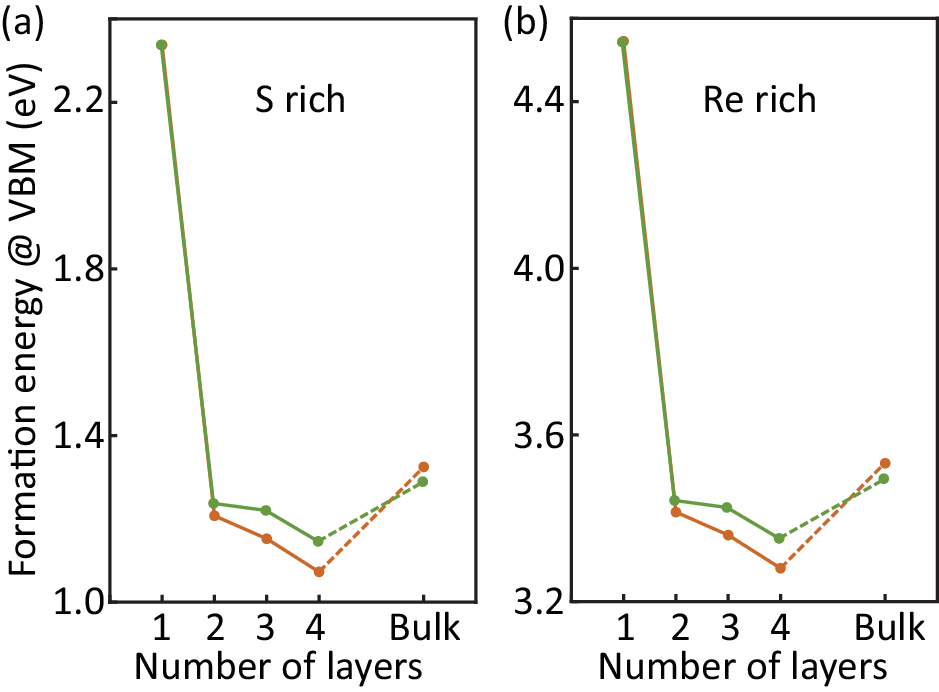}
\end{center}
\caption{Defect formation energy of S$_{Re}$ as function of number of layers from monolayer to bulk ReS$_2$ of AA and AB stacking order under (a) S-rich and (b) Re-rich conditions.}
\label{fig:fig5}
\end{figure*}

\begin{figure*}[htb]
\begin{center}
\includegraphics[width=0.6\textwidth]{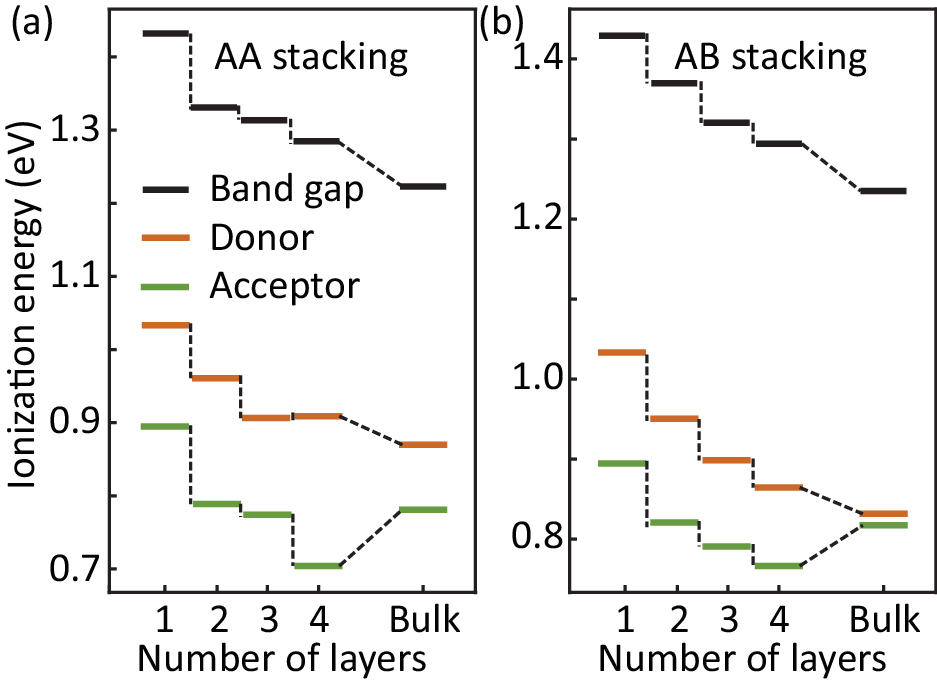}
\end{center}
\caption{Ionization energy of S$_{Re}$ as a function of the number of layers from monolayer to bulk ReS$_2$ of (a) AA and (b)
AB stacking order.}
\label{fig:fig6}
\end{figure*}

\begin{figure*}[t!]
\begin{center}
\includegraphics[width=0.8\textwidth]{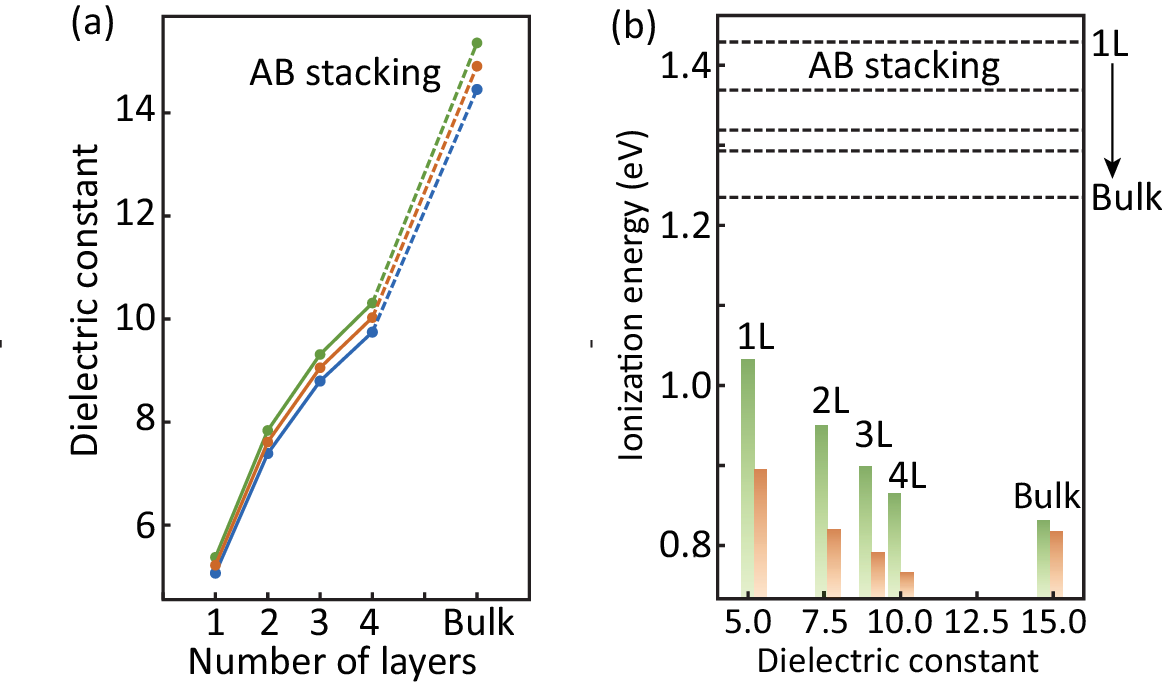}
\end{center}
\caption{Static dielectric constant and defect transition level with respect to dielectric constant for AB stacking. (a) Static dielectric constant along x, y directions and average value as a function of the number of layers (b) Ionization energy of S$_{Re}$ as a function of average static dielectric constant. The green and orange bars represent the ionization energy of acceptors and donors, respectively.}
\label{fig:fig7}
\end{figure*}

\begin{figure*}[htb]
\begin{center}
\includegraphics[width=0.8\textwidth]{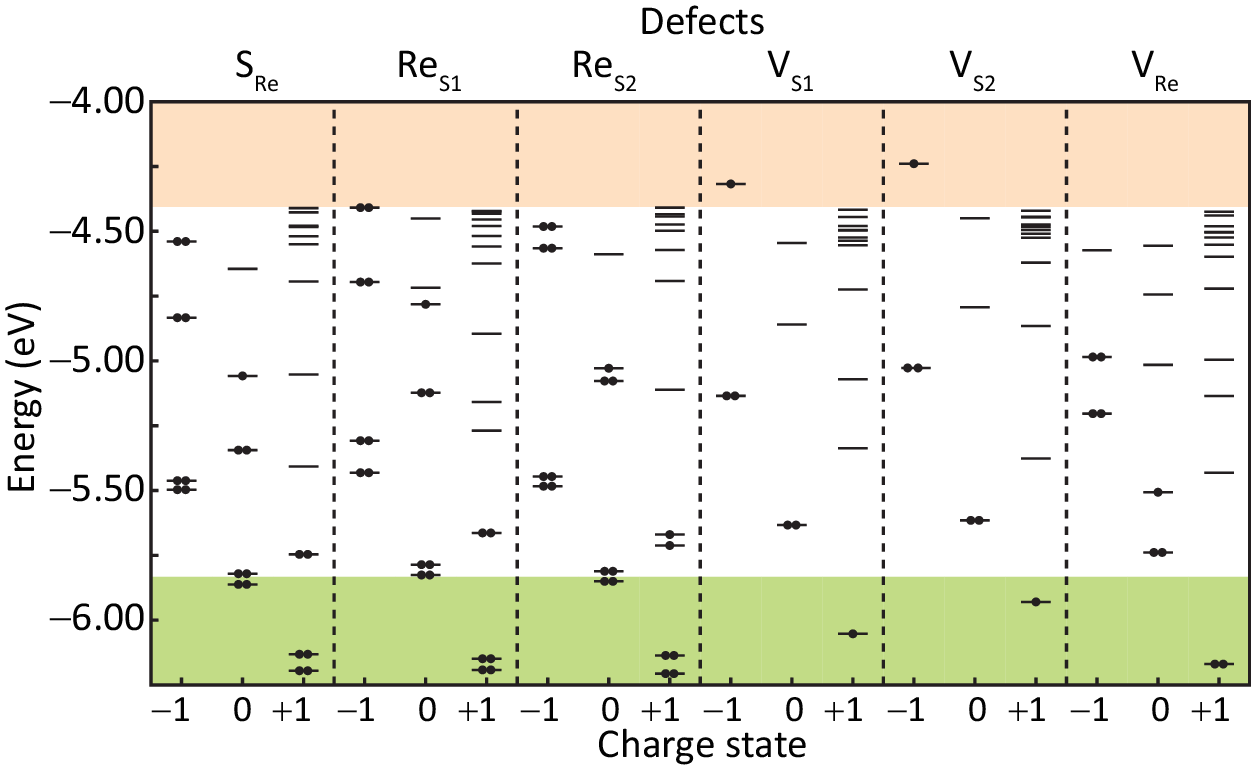}
\end{center}
\caption{Kohn-Sham levels of V$_{S1}$, V$_{S2}$, V$_{Re}$, Re$_{S1}$, Re$_{S2}$ and (f) S$_{Re}$ of monolayer ReS$_2$. For each point defect, the -1, 0 and +1 charge states are divided into three sub-panels. The valence-band (VB) and conduction-band (CB) energy regions are plotted in green and orange, respectively.}
\label{fig:fig8}
\end{figure*}

\begin{figure*}[htb]
\begin{center}
\includegraphics[width=0.8\textwidth]{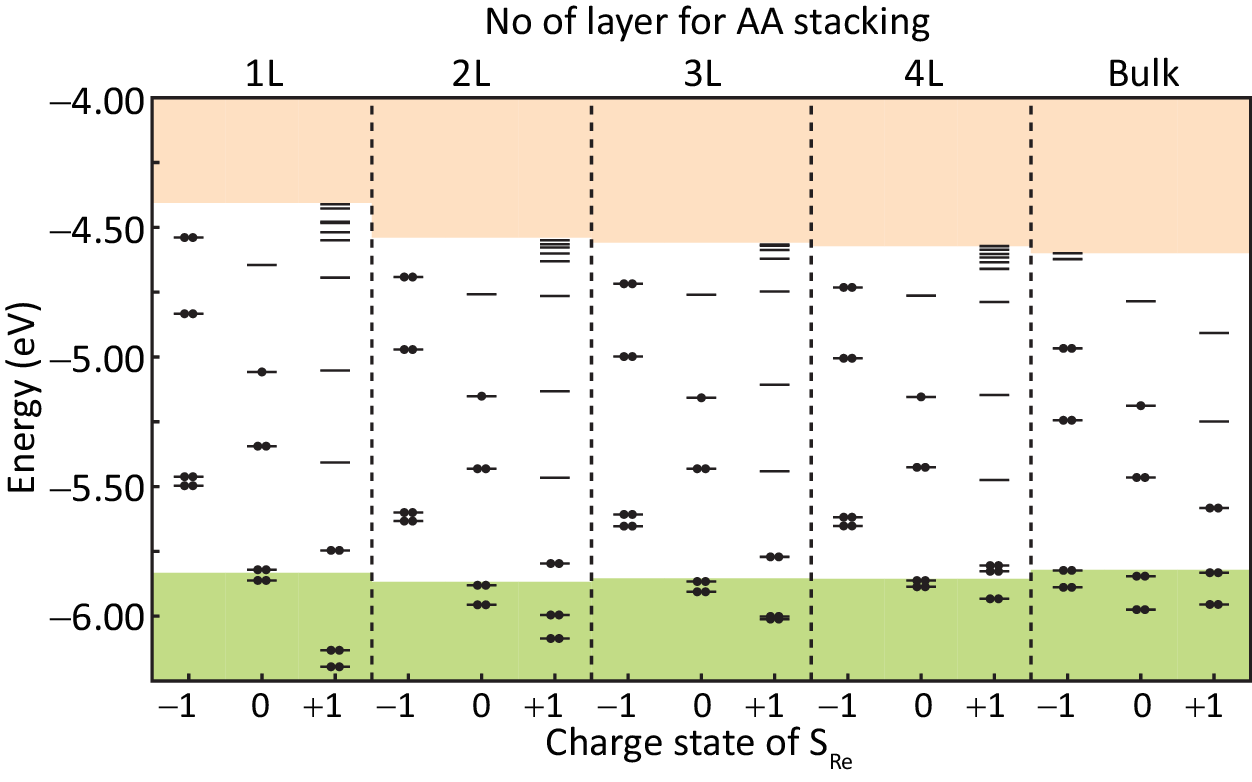}
\end{center}
\caption{Kohn-Sham states of S$_{Re}$ in monolayer to bulk ReS$_2$ of AA stacking. For each point defect, the -1, 0 and +1 charge states are divided into three sub-panels. The valence-band (VB) and conduction-band (CB) energy regions are plotted in green and orange, respectively.}
\label{fig:fig9}
\end{figure*}

\begin{figure*}[htb]
\begin{center}
\includegraphics[width=0.8\textwidth]{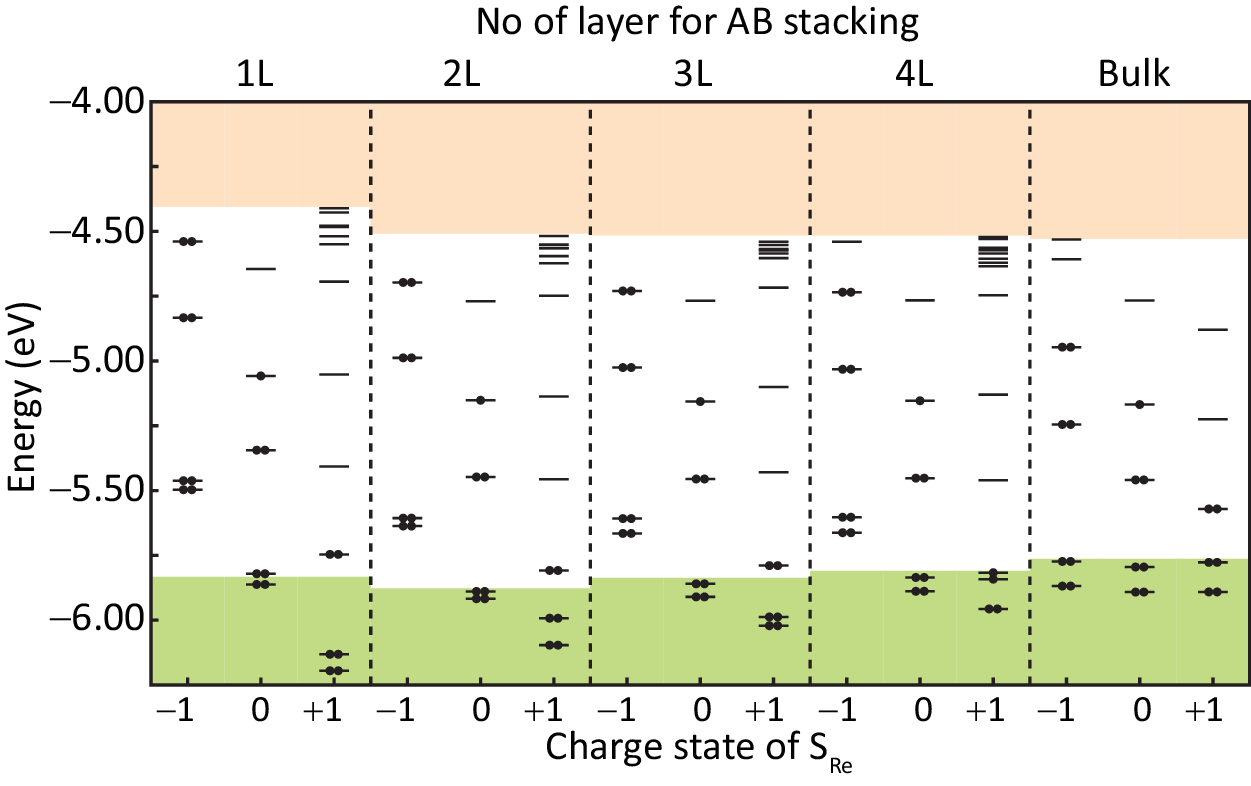}
\end{center}
\caption{Kohn-Sham states of S$_{Re}$ in monolayer to bulk ReS$_2$ of AB stacking. For each point defect, the -1, 0 and +1 charge states are divided into three sub-panels. The valence-band (VB) and conduction-band (CB) energy regions are plotted in green and orange, respectively.}
\label{fig:fig10}
\end{figure*}

\begin{figure*}[htb]
\begin{center}
\includegraphics[width=1.0\textwidth]{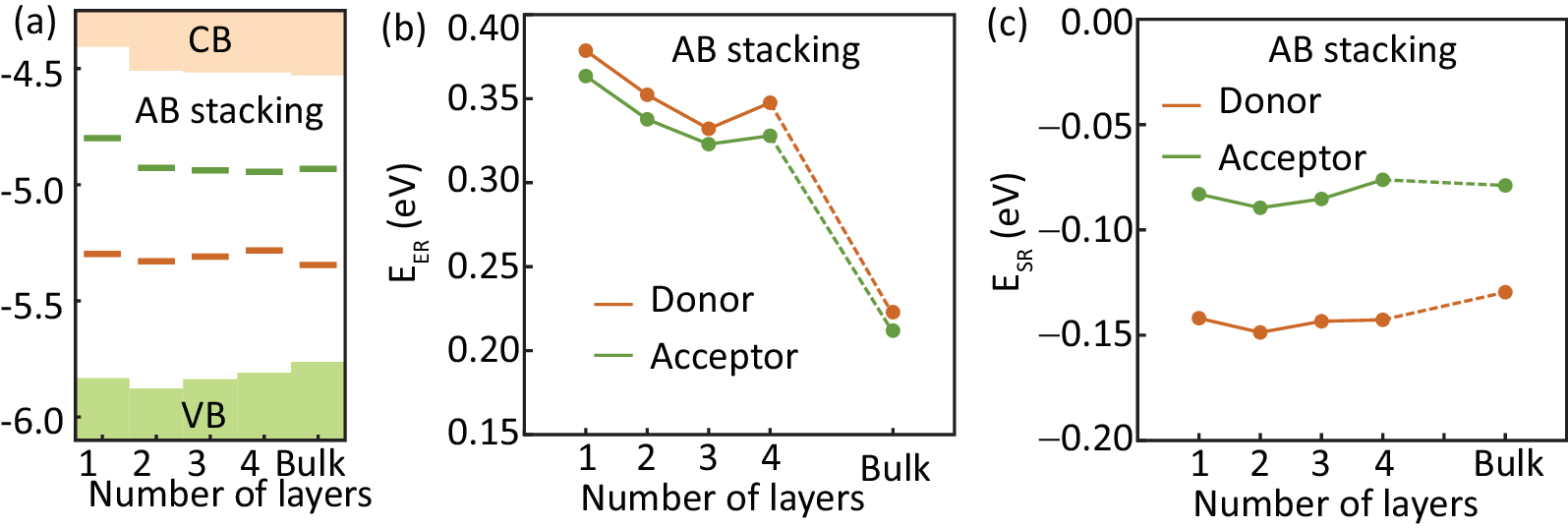}
\end{center}
\caption{(a) Defect transition-energy levels of S$_{Re}$ as function of number of layers for AB stacking. (b) Electronic relaxation energy ($E_{ER}$) and (c) structural relaxation energy ($E_{SR}$) for AB stacking.}
\label{fig:fig11}
\end{figure*}

\begin{figure*}[htb]
\begin{center}
\includegraphics[width=0.5\textwidth]{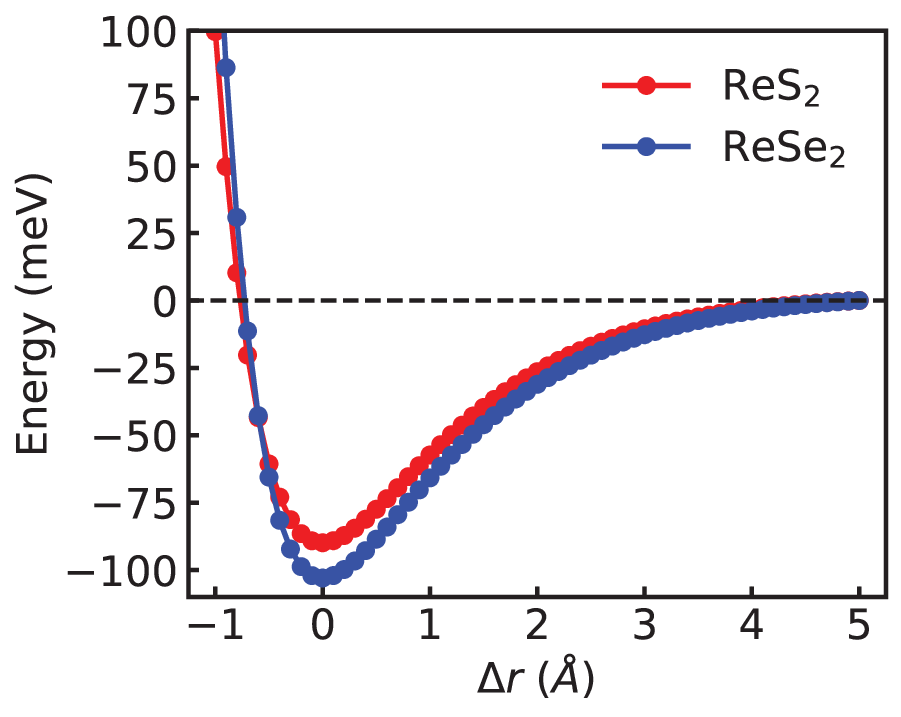}
\end{center}
\caption{Total energy per formula unit of ReS$_2$ and ReSe$_2$ as a function of interlayer separation, illustrating the strength of interlayer coupling.}
\label{fig:fig12}
\end{figure*}

\clearpage

\providecommand{\latin}[1]{#1}
\makeatletter
\providecommand{\doi}
  {\begingroup\let\do\@makeother\dospecials
  \catcode`\{=1 \catcode`\}=2 \doi@aux}
\providecommand{\doi@aux}[1]{\endgroup\texttt{#1}}
\makeatother
\providecommand*\mcitethebibliography{\thebibliography}
\csname @ifundefined\endcsname{endmcitethebibliography}
  {\let\endmcitethebibliography\endthebibliography}{}


\begin{thebibliography}{68}%
\makeatletter
\providecommand \@ifxundefined [1]{%
 \@ifx{#1\undefined}
}%
\providecommand \@ifnum [1]{%
 \ifnum #1\expandafter \@firstoftwo
 \else \expandafter \@secondoftwo
 \fi
}%
\providecommand \@ifx [1]{%
 \ifx #1\expandafter \@firstoftwo
 \else \expandafter \@secondoftwo
 \fi
}%
\providecommand \natexlab [1]{#1}%
\providecommand \enquote  [1]{``#1''}%
\providecommand \bibnamefont  [1]{#1}%
\providecommand \bibfnamefont [1]{#1}%
\providecommand \citenamefont [1]{#1}%
\providecommand \href@noop [0]{\@secondoftwo}%
\providecommand \href [0]{\begingroup \@sanitize@url \@href}%
\providecommand \@href[1]{\@@startlink{#1}\@@href}%
\providecommand \@@href[1]{\endgroup#1\@@endlink}%
\providecommand \@sanitize@url [0]{\catcode `\\12\catcode `\$12\catcode
  `\&12\catcode `\#12\catcode `\^12\catcode `\_12\catcode `\%12\relax}%
\providecommand \@@startlink[1]{}%
\providecommand \@@endlink[0]{}%
\providecommand \url  [0]{\begingroup\@sanitize@url \@url }%
\providecommand \@url [1]{\endgroup\@href {#1}{\urlprefix }}%
\providecommand \urlprefix  [0]{URL }%
\providecommand \Eprint [0]{\href }%
\providecommand \doibase [0]{https://doi.org/}%
\providecommand \selectlanguage [0]{\@gobble}%
\providecommand \bibinfo  [0]{\@secondoftwo}%
\providecommand \bibfield  [0]{\@secondoftwo}%
\providecommand \translation [1]{[#1]}%
\providecommand \BibitemOpen [0]{}%
\providecommand \bibitemStop [0]{}%
\providecommand \bibitemNoStop [0]{.\EOS\space}%
\providecommand \EOS [0]{\spacefactor3000\relax}%
\providecommand \BibitemShut  [1]{\csname bibitem#1\endcsname}%
\let\auto@bib@innerbib\@empty
\bibitem [{\citenamefont {Ko}\ \emph {et~al.}(2024)\citenamefont {Ko},
  \citenamefont {Jang}, \citenamefont {Kwon},\ and\ \citenamefont
  {Suh}}]{Ko2024}%
  \BibitemOpen
  \bibfield  {author} {\bibinfo {author} {\bibfnamefont {K.}~\bibnamefont
  {Ko}}, \bibinfo {author} {\bibfnamefont {M.}~\bibnamefont {Jang}}, \bibinfo
  {author} {\bibfnamefont {J.}~\bibnamefont {Kwon}},\ and\ \bibinfo {author}
  {\bibfnamefont {J.}~\bibnamefont {Suh}},\ }\bibfield  {title} {\bibinfo
  {title} {{Native point defects in 2D transition metal dichalcogenides: A
  perspective bridging intrinsic physical properties and device
  applications}},\ }\href {https://doi.org/10.1063/5.0185604} {\bibfield
  {journal} {\bibinfo  {journal} {J. Appl. Phys.}\ }\textbf {\bibinfo {volume}
  {135}},\ \bibinfo {pages} {100901} (\bibinfo {year} {2024})}\BibitemShut
  {NoStop}%
\bibitem [{\citenamefont {Singh}\ and\ \citenamefont
  {Singh}(2019)}]{Singh2019}%
  \BibitemOpen
  \bibfield  {author} {\bibinfo {author} {\bibfnamefont {A.}~\bibnamefont
  {Singh}}\ and\ \bibinfo {author} {\bibfnamefont {A.~K.}\ \bibnamefont
  {Singh}},\ }\bibfield  {title} {\bibinfo {title} {{Origin of n-type
  conductivity of monolayer MoS$_2$}},\ }\href
  {https://doi.org/10.1103/physrevb.99.121201} {\bibfield  {journal} {\bibinfo
  {journal} {Phys. Rev. B}\ }\textbf {\bibinfo {volume} {99}},\ \bibinfo
  {pages} {121201(R)} (\bibinfo {year} {2019})}\BibitemShut {NoStop}%
\bibitem [{\citenamefont {Jiang}\ \emph {et~al.}(2018)\citenamefont {Jiang},
  \citenamefont {Ling}, \citenamefont {Xu}, \citenamefont {Wang}, \citenamefont
  {Niu}, \citenamefont {Zafar}, \citenamefont {Yan}, \citenamefont {Wang},
  \citenamefont {You}, \citenamefont {Sun}, \citenamefont {Lu}, \citenamefont
  {Wang},\ and\ \citenamefont {Ni}}]{Jiang2018}%
  \BibitemOpen
  \bibfield  {author} {\bibinfo {author} {\bibfnamefont {J.}~\bibnamefont
  {Jiang}}, \bibinfo {author} {\bibfnamefont {C.}~\bibnamefont {Ling}},
  \bibinfo {author} {\bibfnamefont {T.}~\bibnamefont {Xu}}, \bibinfo {author}
  {\bibfnamefont {W.}~\bibnamefont {Wang}}, \bibinfo {author} {\bibfnamefont
  {X.}~\bibnamefont {Niu}}, \bibinfo {author} {\bibfnamefont {A.}~\bibnamefont
  {Zafar}}, \bibinfo {author} {\bibfnamefont {Z.}~\bibnamefont {Yan}}, \bibinfo
  {author} {\bibfnamefont {X.}~\bibnamefont {Wang}}, \bibinfo {author}
  {\bibfnamefont {Y.}~\bibnamefont {You}}, \bibinfo {author} {\bibfnamefont
  {L.}~\bibnamefont {Sun}}, \bibinfo {author} {\bibfnamefont {J.}~\bibnamefont
  {Lu}}, \bibinfo {author} {\bibfnamefont {J.}~\bibnamefont {Wang}},\ and\
  \bibinfo {author} {\bibfnamefont {Z.}~\bibnamefont {Ni}},\ }\bibfield
  {title} {\bibinfo {title} {{Defect Engineering for Modulating the Trap States
  in 2D Photoconductors}},\ }\href {https://doi.org/10.1002/adma.201804332}
  {\bibfield  {journal} {\bibinfo  {journal} {Adv. Mater.}\ }\textbf {\bibinfo
  {volume} {30}},\ \bibinfo {pages} {1804332} (\bibinfo {year}
  {2018})}\BibitemShut {NoStop}%
\bibitem [{\citenamefont {Chen}\ \emph {et~al.}(2024)\citenamefont {Chen},
  \citenamefont {Hsu}, \citenamefont {Liang}, \citenamefont {Wu}, \citenamefont
  {Chen}, \citenamefont {Huang}, \citenamefont {Li}, \citenamefont {Radu},\
  and\ \citenamefont {Chiu}}]{Chen2024}%
  \BibitemOpen
  \bibfield  {author} {\bibinfo {author} {\bibfnamefont {H.-Y.}\ \bibnamefont
  {Chen}}, \bibinfo {author} {\bibfnamefont {H.-C.}\ \bibnamefont {Hsu}},
  \bibinfo {author} {\bibfnamefont {J.-Y.}\ \bibnamefont {Liang}}, \bibinfo
  {author} {\bibfnamefont {B.-H.}\ \bibnamefont {Wu}}, \bibinfo {author}
  {\bibfnamefont {Y.-F.}\ \bibnamefont {Chen}}, \bibinfo {author}
  {\bibfnamefont {C.-C.}\ \bibnamefont {Huang}}, \bibinfo {author}
  {\bibfnamefont {M.-Y.}\ \bibnamefont {Li}}, \bibinfo {author} {\bibfnamefont
  {I.~P.}\ \bibnamefont {Radu}},\ and\ \bibinfo {author} {\bibfnamefont
  {Y.-P.}\ \bibnamefont {Chiu}},\ }\bibfield  {title} {\bibinfo {title}
  {{Atomically Resolved Defect-Engineering Scattering Potential in 2D
  Semiconductors}},\ }\href {https://doi.org/10.1021/acsnano.4c02066}
  {\bibfield  {journal} {\bibinfo  {journal} {ACS Nano}\ }\textbf {\bibinfo
  {volume} {18}},\ \bibinfo {pages} {17622–17629} (\bibinfo {year}
  {2024})}\BibitemShut {NoStop}%
\bibitem [{\citenamefont {Zhang}\ \emph {et~al.}(2019)\citenamefont {Zhang},
  \citenamefont {Chu}, \citenamefont {Zheng}, \citenamefont {Benderskii},
  \citenamefont {Prezhdo},\ and\ \citenamefont {Zhao}}]{Zhang2019}%
  \BibitemOpen
  \bibfield  {author} {\bibinfo {author} {\bibfnamefont {L.}~\bibnamefont
  {Zhang}}, \bibinfo {author} {\bibfnamefont {W.}~\bibnamefont {Chu}}, \bibinfo
  {author} {\bibfnamefont {Q.}~\bibnamefont {Zheng}}, \bibinfo {author}
  {\bibfnamefont {A.~V.}\ \bibnamefont {Benderskii}}, \bibinfo {author}
  {\bibfnamefont {O.~V.}\ \bibnamefont {Prezhdo}},\ and\ \bibinfo {author}
  {\bibfnamefont {J.}~\bibnamefont {Zhao}},\ }\bibfield  {title} {\bibinfo
  {title} {{Suppression of Electron–Hole Recombination by Intrinsic Defects
  in 2D Monoelemental Material}},\ }\href
  {https://doi.org/10.1021/acs.jpclett.9b02620} {\bibfield  {journal} {\bibinfo
   {journal} {J. Phys. Chem. Lett.}\ }\textbf {\bibinfo {volume} {10}},\
  \bibinfo {pages} {6151–6158} (\bibinfo {year} {2019})}\BibitemShut
  {NoStop}%
\bibitem [{\citenamefont {Turunen}\ \emph {et~al.}(2022)\citenamefont
  {Turunen}, \citenamefont {Brotons-Gisbert}, \citenamefont {Dai},
  \citenamefont {Wang}, \citenamefont {Scerri}, \citenamefont {Bonato},
  \citenamefont {J\"{o}ns}, \citenamefont {Sun},\ and\ \citenamefont
  {Gerardot}}]{Turunen2022}%
  \BibitemOpen
  \bibfield  {author} {\bibinfo {author} {\bibfnamefont {M.}~\bibnamefont
  {Turunen}}, \bibinfo {author} {\bibfnamefont {M.}~\bibnamefont
  {Brotons-Gisbert}}, \bibinfo {author} {\bibfnamefont {Y.}~\bibnamefont
  {Dai}}, \bibinfo {author} {\bibfnamefont {Y.}~\bibnamefont {Wang}}, \bibinfo
  {author} {\bibfnamefont {E.}~\bibnamefont {Scerri}}, \bibinfo {author}
  {\bibfnamefont {C.}~\bibnamefont {Bonato}}, \bibinfo {author} {\bibfnamefont
  {K.~D.}\ \bibnamefont {J\"{o}ns}}, \bibinfo {author} {\bibfnamefont
  {Z.}~\bibnamefont {Sun}},\ and\ \bibinfo {author} {\bibfnamefont {B.~D.}\
  \bibnamefont {Gerardot}},\ }\bibfield  {title} {\bibinfo {title} {{Quantum
  photonics with layered 2D materials}},\ }\href
  {https://doi.org/10.1038/s42254-021-00408-0} {\bibfield  {journal} {\bibinfo
  {journal} {Nat. Rev. Phys.}\ }\textbf {\bibinfo {volume} {4}},\ \bibinfo
  {pages} {219–236} (\bibinfo {year} {2022})}\BibitemShut {NoStop}%
\bibitem [{\citenamefont {Dey}\ \emph {et~al.}(2025)\citenamefont {Dey},
  \citenamefont {Meher},\ and\ \citenamefont {Singh}}]{Dey2025}%
  \BibitemOpen
  \bibfield  {author} {\bibinfo {author} {\bibfnamefont {M.}~\bibnamefont
  {Dey}}, \bibinfo {author} {\bibfnamefont {S.}~\bibnamefont {Meher}},\ and\
  \bibinfo {author} {\bibfnamefont {A.~K.}\ \bibnamefont {Singh}},\ }\bibfield
  {title} {\bibinfo {title} {{Carbon with Stone-Wales defect as quantum emitter
  in h-BN}},\ }\href {https://doi.org/10.1103/physrevb.111.104109} {\bibfield
  {journal} {\bibinfo  {journal} {Phys. Rev. B}\ }\textbf {\bibinfo {volume}
  {111}},\ \bibinfo {pages} {104109} (\bibinfo {year} {2025})}\BibitemShut
  {NoStop}%
\bibitem [{\citenamefont {Arora}\ \emph {et~al.}(2021)\citenamefont {Arora},
  \citenamefont {Nayak}, \citenamefont {Bhattacharyya}, \citenamefont {Maity},
  \citenamefont {Singh}, \citenamefont {Krishnan},\ and\ \citenamefont
  {Rao}}]{Arora2021}%
  \BibitemOpen
  \bibfield  {author} {\bibinfo {author} {\bibfnamefont {A.}~\bibnamefont
  {Arora}}, \bibinfo {author} {\bibfnamefont {P.~K.}\ \bibnamefont {Nayak}},
  \bibinfo {author} {\bibfnamefont {S.}~\bibnamefont {Bhattacharyya}}, \bibinfo
  {author} {\bibfnamefont {N.}~\bibnamefont {Maity}}, \bibinfo {author}
  {\bibfnamefont {A.~K.}\ \bibnamefont {Singh}}, \bibinfo {author}
  {\bibfnamefont {A.}~\bibnamefont {Krishnan}},\ and\ \bibinfo {author}
  {\bibfnamefont {M.~S.~R.}\ \bibnamefont {Rao}},\ }\bibfield  {title}
  {\bibinfo {title} {{Interlayer excitonic states in MoSe$_2$/MoS$_2$ van der
  Waals heterostructures}},\ }\href
  {https://doi.org/10.1103/physrevb.103.205406} {\bibfield  {journal} {\bibinfo
   {journal} {Phys. Rev. B}\ }\textbf {\bibinfo {volume} {103}},\ \bibinfo
  {pages} {205406} (\bibinfo {year} {2021})}\BibitemShut {NoStop}%
\bibitem [{\citenamefont {Mandal}\ \emph {et~al.}(2023)\citenamefont {Mandal},
  \citenamefont {Maity}, \citenamefont {Barman}, \citenamefont {Srivastava},
  \citenamefont {Singh}, \citenamefont {Nayak},\ and\ \citenamefont
  {Sethupathi}}]{Mandal2023}%
  \BibitemOpen
  \bibfield  {author} {\bibinfo {author} {\bibfnamefont {M.}~\bibnamefont
  {Mandal}}, \bibinfo {author} {\bibfnamefont {N.}~\bibnamefont {Maity}},
  \bibinfo {author} {\bibfnamefont {P.~K.}\ \bibnamefont {Barman}}, \bibinfo
  {author} {\bibfnamefont {A.}~\bibnamefont {Srivastava}}, \bibinfo {author}
  {\bibfnamefont {A.~K.}\ \bibnamefont {Singh}}, \bibinfo {author}
  {\bibfnamefont {P.~K.}\ \bibnamefont {Nayak}},\ and\ \bibinfo {author}
  {\bibfnamefont {K.}~\bibnamefont {Sethupathi}},\ }\bibfield  {title}
  {\bibinfo {title} {{Probing angle-dependent thermal conductivity in twisted
  bilayer {MoSe$_2$}}},\ }\href {https://doi.org/10.1103/physrevb.108.115439}
  {\bibfield  {journal} {\bibinfo  {journal} {Phys. Rev. B}\ }\textbf {\bibinfo
  {volume} {108}},\ \bibinfo {pages} {115439} (\bibinfo {year}
  {2023})}\BibitemShut {NoStop}%
\bibitem [{\citenamefont {Mishra}\ \emph {et~al.}(2024)\citenamefont {Mishra},
  \citenamefont {Maity},\ and\ \citenamefont {Singh}}]{Mishra2024}%
  \BibitemOpen
  \bibfield  {author} {\bibinfo {author} {\bibfnamefont {S.}~\bibnamefont
  {Mishra}}, \bibinfo {author} {\bibfnamefont {N.}~\bibnamefont {Maity}},\ and\
  \bibinfo {author} {\bibfnamefont {A.~K.}\ \bibnamefont {Singh}},\ }\bibfield
  {title} {\bibinfo {title} {{Symmetry-assisted anomalous Hall conductivity in
  a CrS$_2$-CrBr$_3$ heterostructure}},\ }\href
  {https://doi.org/10.1103/physrevb.110.125406} {\bibfield  {journal} {\bibinfo
   {journal} {Phys. Rev. B}\ }\textbf {\bibinfo {volume} {110}},\ \bibinfo
  {pages} {125406} (\bibinfo {year} {2024})}\BibitemShut {NoStop}%
\bibitem [{\citenamefont {Mak}\ \emph {et~al.}(2010)\citenamefont {Mak},
  \citenamefont {Lee}, \citenamefont {Hone}, \citenamefont {Shan},\ and\
  \citenamefont {Heinz}}]{Mak2010}%
  \BibitemOpen
  \bibfield  {author} {\bibinfo {author} {\bibfnamefont {K.~F.}\ \bibnamefont
  {Mak}}, \bibinfo {author} {\bibfnamefont {C.}~\bibnamefont {Lee}}, \bibinfo
  {author} {\bibfnamefont {J.}~\bibnamefont {Hone}}, \bibinfo {author}
  {\bibfnamefont {J.}~\bibnamefont {Shan}},\ and\ \bibinfo {author}
  {\bibfnamefont {T.~F.}\ \bibnamefont {Heinz}},\ }\bibfield  {title} {\bibinfo
  {title} {{Atomically thin MoS$_2$: a new direct-gap semiconductor}},\ }\href
  {https://doi.org/10.1103/physrevlett.105.136805} {\bibfield  {journal}
  {\bibinfo  {journal} {Phys. Rev. Lett.}\ }\textbf {\bibinfo {volume} {105}},\
  \bibinfo {pages} {136805} (\bibinfo {year} {2010})}\BibitemShut {NoStop}%
\bibitem [{\citenamefont {Wang}\ \emph {et~al.}(2012)\citenamefont {Wang},
  \citenamefont {Kalantar-Zadeh}, \citenamefont {Kis}, \citenamefont
  {Coleman},\ and\ \citenamefont {Strano}}]{Wang2012}%
  \BibitemOpen
  \bibfield  {author} {\bibinfo {author} {\bibfnamefont {Q.~H.}\ \bibnamefont
  {Wang}}, \bibinfo {author} {\bibfnamefont {K.}~\bibnamefont
  {Kalantar-Zadeh}}, \bibinfo {author} {\bibfnamefont {A.}~\bibnamefont {Kis}},
  \bibinfo {author} {\bibfnamefont {J.~N.}\ \bibnamefont {Coleman}},\ and\
  \bibinfo {author} {\bibfnamefont {M.~S.}\ \bibnamefont {Strano}},\ }\bibfield
   {title} {\bibinfo {title} {Electronics and optoelectronics of
  two-dimensional transition metal dichalcogenides},\ }\href
  {https://doi.org/10.1038/nnano.2012.193} {\bibfield  {journal} {\bibinfo
  {journal} {Nat. Nanotechnol.}\ }\textbf {\bibinfo {volume} {7}},\ \bibinfo
  {pages} {699–712} (\bibinfo {year} {2012})}\BibitemShut {NoStop}%
\bibitem [{\citenamefont {Radisavljevic}\ \emph {et~al.}(2011)\citenamefont
  {Radisavljevic}, \citenamefont {Radenovic}, \citenamefont {Brivio},
  \citenamefont {Giacometti},\ and\ \citenamefont {Kis}}]{Radisavljevic2011}%
  \BibitemOpen
  \bibfield  {author} {\bibinfo {author} {\bibfnamefont {B.}~\bibnamefont
  {Radisavljevic}}, \bibinfo {author} {\bibfnamefont {A.}~\bibnamefont
  {Radenovic}}, \bibinfo {author} {\bibfnamefont {J.}~\bibnamefont {Brivio}},
  \bibinfo {author} {\bibfnamefont {V.}~\bibnamefont {Giacometti}},\ and\
  \bibinfo {author} {\bibfnamefont {A.}~\bibnamefont {Kis}},\ }\bibfield
  {title} {\bibinfo {title} {{Single-layer MoS$_2$ transistors}},\ }\href
  {https://doi.org/10.1038/nnano.2010.279} {\bibfield  {journal} {\bibinfo
  {journal} {Nat. Nanotechnol.}\ }\textbf {\bibinfo {volume} {6}},\ \bibinfo
  {pages} {147} (\bibinfo {year} {2011})}\BibitemShut {NoStop}%
\bibitem [{\citenamefont {Chhowalla}\ \emph {et~al.}(2013)\citenamefont
  {Chhowalla}, \citenamefont {Shin}, \citenamefont {Eda}, \citenamefont {Li},
  \citenamefont {Loh},\ and\ \citenamefont {Zhang}}]{Chhowalla2013}%
  \BibitemOpen
  \bibfield  {author} {\bibinfo {author} {\bibfnamefont {M.}~\bibnamefont
  {Chhowalla}}, \bibinfo {author} {\bibfnamefont {H.~S.}\ \bibnamefont {Shin}},
  \bibinfo {author} {\bibfnamefont {G.}~\bibnamefont {Eda}}, \bibinfo {author}
  {\bibfnamefont {L.-J.}\ \bibnamefont {Li}}, \bibinfo {author} {\bibfnamefont
  {K.~P.}\ \bibnamefont {Loh}},\ and\ \bibinfo {author} {\bibfnamefont
  {H.}~\bibnamefont {Zhang}},\ }\bibfield  {title} {\bibinfo {title} {The
  chemistry of two-dimensional layered transition metal dichalcogenide
  nanosheets},\ }\href {https://doi.org/10.1038/nchem.1589} {\bibfield
  {journal} {\bibinfo  {journal} {Nat. Chem.}\ }\textbf {\bibinfo {volume}
  {5}},\ \bibinfo {pages} {263–275} (\bibinfo {year} {2013})}\BibitemShut
  {NoStop}%
\bibitem [{\citenamefont {Bhattacharyya}\ and\ \citenamefont
  {Singh}(2012)}]{Bhattacharyya2012}%
  \BibitemOpen
  \bibfield  {author} {\bibinfo {author} {\bibfnamefont {S.}~\bibnamefont
  {Bhattacharyya}}\ and\ \bibinfo {author} {\bibfnamefont {A.~K.}\ \bibnamefont
  {Singh}},\ }\bibfield  {title} {\bibinfo {title} {Semiconductor-metal
  transition in semiconducting bilayer sheets of transition-metal
  dichalcogenides},\ }\href {https://doi.org/10.1103/physrevb.86.075454}
  {\bibfield  {journal} {\bibinfo  {journal} {Phys. Rev. B}\ }\textbf {\bibinfo
  {volume} {86}},\ \bibinfo {pages} {075454} (\bibinfo {year}
  {2012})}\BibitemShut {NoStop}%
\bibitem [{\citenamefont {Tongay}\ \emph {et~al.}(2012)\citenamefont {Tongay},
  \citenamefont {Zhou}, \citenamefont {Ataca}, \citenamefont {Lo},
  \citenamefont {Matthews}, \citenamefont {Li}, \citenamefont {Grossman},\ and\
  \citenamefont {Wu}}]{Tongay2012}%
  \BibitemOpen
  \bibfield  {author} {\bibinfo {author} {\bibfnamefont {S.}~\bibnamefont
  {Tongay}}, \bibinfo {author} {\bibfnamefont {J.}~\bibnamefont {Zhou}},
  \bibinfo {author} {\bibfnamefont {C.}~\bibnamefont {Ataca}}, \bibinfo
  {author} {\bibfnamefont {K.}~\bibnamefont {Lo}}, \bibinfo {author}
  {\bibfnamefont {T.~S.}\ \bibnamefont {Matthews}}, \bibinfo {author}
  {\bibfnamefont {J.}~\bibnamefont {Li}}, \bibinfo {author} {\bibfnamefont
  {J.~C.}\ \bibnamefont {Grossman}},\ and\ \bibinfo {author} {\bibfnamefont
  {J.}~\bibnamefont {Wu}},\ }\bibfield  {title} {\bibinfo {title} {{Thermally
  driven crossover from indirect toward direct bandgap in 2D semiconductors:
  MoSe$_2$ versus MoS$_2$}},\ }\href {https://doi.org/10.1021/nl302584w}
  {\bibfield  {journal} {\bibinfo  {journal} {Nano Lett.}\ }\textbf {\bibinfo
  {volume} {12}},\ \bibinfo {pages} {5576} (\bibinfo {year}
  {2012})}\BibitemShut {NoStop}%
\bibitem [{\citenamefont {Splendiani}\ \emph {et~al.}(2010)\citenamefont
  {Splendiani}, \citenamefont {Sun}, \citenamefont {Zhang}, \citenamefont {Li},
  \citenamefont {Kim}, \citenamefont {Chim}, \citenamefont {Galli},\ and\
  \citenamefont {Wang}}]{Splendiani2010}%
  \BibitemOpen
  \bibfield  {author} {\bibinfo {author} {\bibfnamefont {A.}~\bibnamefont
  {Splendiani}}, \bibinfo {author} {\bibfnamefont {L.}~\bibnamefont {Sun}},
  \bibinfo {author} {\bibfnamefont {Y.}~\bibnamefont {Zhang}}, \bibinfo
  {author} {\bibfnamefont {T.}~\bibnamefont {Li}}, \bibinfo {author}
  {\bibfnamefont {J.}~\bibnamefont {Kim}}, \bibinfo {author} {\bibfnamefont
  {C.-Y.}\ \bibnamefont {Chim}}, \bibinfo {author} {\bibfnamefont
  {G.}~\bibnamefont {Galli}},\ and\ \bibinfo {author} {\bibfnamefont
  {F.}~\bibnamefont {Wang}},\ }\bibfield  {title} {\bibinfo {title} {{Emerging
  photoluminescence in monolayer MoS$_2$}},\ }\href
  {https://doi.org/10.1021/nl903868w} {\bibfield  {journal} {\bibinfo
  {journal} {Nano Lett.}\ }\textbf {\bibinfo {volume} {10}},\ \bibinfo {pages}
  {1271} (\bibinfo {year} {2010})}\BibitemShut {NoStop}%
\bibitem [{\citenamefont {Qiao}\ \emph {et~al.}(2014)\citenamefont {Qiao},
  \citenamefont {Kong}, \citenamefont {Hu}, \citenamefont {Yang},\ and\
  \citenamefont {Ji}}]{Qiao2014}%
  \BibitemOpen
  \bibfield  {author} {\bibinfo {author} {\bibfnamefont {J.}~\bibnamefont
  {Qiao}}, \bibinfo {author} {\bibfnamefont {X.}~\bibnamefont {Kong}}, \bibinfo
  {author} {\bibfnamefont {Z.-X.}\ \bibnamefont {Hu}}, \bibinfo {author}
  {\bibfnamefont {F.}~\bibnamefont {Yang}},\ and\ \bibinfo {author}
  {\bibfnamefont {W.}~\bibnamefont {Ji}},\ }\bibfield  {title} {\bibinfo
  {title} {{High-mobility transport anisotropy and linear dichroism in
  few-layer black phosphorus}},\ }\href {https://doi.org/10.1038/ncomms5475}
  {\bibfield  {journal} {\bibinfo  {journal} {Nat. Commun.}\ }\textbf {\bibinfo
  {volume} {5}},\ \bibinfo {pages} {4475} (\bibinfo {year} {2014})}\BibitemShut
  {NoStop}%
\bibitem [{\citenamefont {Zhao}\ and\ \citenamefont {Liu}(2018)}]{Zhao2018}%
  \BibitemOpen
  \bibfield  {author} {\bibinfo {author} {\bibfnamefont {Z.-Y.}\ \bibnamefont
  {Zhao}}\ and\ \bibinfo {author} {\bibfnamefont {Q.-L.}\ \bibnamefont {Liu}},\
  }\bibfield  {title} {\bibinfo {title} {{Study of the layer-dependent
  properties of MoS$_2$ nanosheets with different crystal structures by DFT
  calculations}},\ }\href {https://doi.org/10.1039/c7cy02252b} {\bibfield
  {journal} {\bibinfo  {journal} {Catal. Sci. Technol.}\ }\textbf {\bibinfo
  {volume} {8}},\ \bibinfo {pages} {1867–1879} (\bibinfo {year}
  {2018})}\BibitemShut {NoStop}%
\bibitem [{\citenamefont {Kayyalha}\ \emph {et~al.}(2016)\citenamefont
  {Kayyalha}, \citenamefont {Maassen}, \citenamefont {Lundstrom}, \citenamefont
  {Shi},\ and\ \citenamefont {Chen}}]{Kayyalha2016}%
  \BibitemOpen
  \bibfield  {author} {\bibinfo {author} {\bibfnamefont {M.}~\bibnamefont
  {Kayyalha}}, \bibinfo {author} {\bibfnamefont {J.}~\bibnamefont {Maassen}},
  \bibinfo {author} {\bibfnamefont {M.}~\bibnamefont {Lundstrom}}, \bibinfo
  {author} {\bibfnamefont {L.}~\bibnamefont {Shi}},\ and\ \bibinfo {author}
  {\bibfnamefont {Y.~P.}\ \bibnamefont {Chen}},\ }\bibfield  {title} {\bibinfo
  {title} {{Gate-tunable and thickness-dependent electronic and thermoelectric
  transport in few-layer MoS$_2$}},\ }\href {https://doi.org/10.1063/1.4963364}
  {\bibfield  {journal} {\bibinfo  {journal} {J. Appl. Phys.}\ }\textbf
  {\bibinfo {volume} {120}},\ \bibinfo {pages} {134305} (\bibinfo {year}
  {2016})}\BibitemShut {NoStop}%
\bibitem [{\citenamefont {Choi}\ \emph {et~al.}(2017)\citenamefont {Choi},
  \citenamefont {Choudhary}, \citenamefont {Han}, \citenamefont {Park},
  \citenamefont {Akinwande},\ and\ \citenamefont {Lee}}]{Choi2017}%
  \BibitemOpen
  \bibfield  {author} {\bibinfo {author} {\bibfnamefont {W.}~\bibnamefont
  {Choi}}, \bibinfo {author} {\bibfnamefont {N.}~\bibnamefont {Choudhary}},
  \bibinfo {author} {\bibfnamefont {G.~H.}\ \bibnamefont {Han}}, \bibinfo
  {author} {\bibfnamefont {J.}~\bibnamefont {Park}}, \bibinfo {author}
  {\bibfnamefont {D.}~\bibnamefont {Akinwande}},\ and\ \bibinfo {author}
  {\bibfnamefont {Y.~H.}\ \bibnamefont {Lee}},\ }\bibfield  {title} {\bibinfo
  {title} {Recent development of two-dimensional transition metal
  dichalcogenides and their applications},\ }\href
  {https://doi.org/10.1016/j.mattod.2016.10.002} {\bibfield  {journal}
  {\bibinfo  {journal} {Mater. Today}\ }\textbf {\bibinfo {volume} {20}},\
  \bibinfo {pages} {116–130} (\bibinfo {year} {2017})}\BibitemShut {NoStop}%
\bibitem [{\citenamefont {Jiao}\ \emph {et~al.}(2019)\citenamefont {Jiao},
  \citenamefont {Jie}, \citenamefont {Yang}, \citenamefont {Wang},
  \citenamefont {Chen}, \citenamefont {Zhang}, \citenamefont {Tang},
  \citenamefont {Wu},\ and\ \citenamefont {Hao}}]{Jiao2019}%
  \BibitemOpen
  \bibfield  {author} {\bibinfo {author} {\bibfnamefont {L.}~\bibnamefont
  {Jiao}}, \bibinfo {author} {\bibfnamefont {W.}~\bibnamefont {Jie}}, \bibinfo
  {author} {\bibfnamefont {Z.}~\bibnamefont {Yang}}, \bibinfo {author}
  {\bibfnamefont {Y.}~\bibnamefont {Wang}}, \bibinfo {author} {\bibfnamefont
  {Z.}~\bibnamefont {Chen}}, \bibinfo {author} {\bibfnamefont {X.}~\bibnamefont
  {Zhang}}, \bibinfo {author} {\bibfnamefont {W.}~\bibnamefont {Tang}},
  \bibinfo {author} {\bibfnamefont {Z.}~\bibnamefont {Wu}},\ and\ \bibinfo
  {author} {\bibfnamefont {J.}~\bibnamefont {Hao}},\ }\bibfield  {title}
  {\bibinfo {title} {{Layer-dependent photoresponse of 2D Mo$_2$ films prepared
  by pulsed laser deposition}},\ }\href {https://doi.org/10.1039/c8tc04612c}
  {\bibfield  {journal} {\bibinfo  {journal} {J. Mater. Chem. C}\ }\textbf
  {\bibinfo {volume} {7}},\ \bibinfo {pages} {2522–2529} (\bibinfo {year}
  {2019})}\BibitemShut {NoStop}%
\bibitem [{\citenamefont {Singh}\ \emph {et~al.}(2022)\citenamefont {Singh},
  \citenamefont {Dey},\ and\ \citenamefont {Singh}}]{Singh2022}%
  \BibitemOpen
  \bibfield  {author} {\bibinfo {author} {\bibfnamefont {A.}~\bibnamefont
  {Singh}}, \bibinfo {author} {\bibfnamefont {M.}~\bibnamefont {Dey}},\ and\
  \bibinfo {author} {\bibfnamefont {A.~K.}\ \bibnamefont {Singh}},\ }\bibfield
  {title} {\bibinfo {title} {Origin of layer-dependent electrical conductivity
  of transition metal dichalcogenides},\ }\href
  {https://doi.org/10.1103/physrevb.105.165430} {\bibfield  {journal} {\bibinfo
   {journal} {Phys. Rev. B}\ }\textbf {\bibinfo {volume} {105}},\ \bibinfo
  {pages} {165430} (\bibinfo {year} {2022})}\BibitemShut {NoStop}%
\bibitem [{\citenamefont {Zhu}\ \emph {et~al.}(2021)\citenamefont {Zhu},
  \citenamefont {Xu}, \citenamefont {Gong}, \citenamefont {Yang},\ and\
  \citenamefont {Yakobson}}]{Zhu2021}%
  \BibitemOpen
  \bibfield  {author} {\bibinfo {author} {\bibfnamefont {G.-J.}\ \bibnamefont
  {Zhu}}, \bibinfo {author} {\bibfnamefont {Y.-G.}\ \bibnamefont {Xu}},
  \bibinfo {author} {\bibfnamefont {X.-G.}\ \bibnamefont {Gong}}, \bibinfo
  {author} {\bibfnamefont {J.-H.}\ \bibnamefont {Yang}},\ and\ \bibinfo
  {author} {\bibfnamefont {B.~I.}\ \bibnamefont {Yakobson}},\ }\bibfield
  {title} {\bibinfo {title} {{Dimensionality-Inhibited Chemical Doping in
  Two-Dimensional Semiconductors: The Phosphorene and MoS$_2$ from
  Charge-Correction Method}},\ }\href
  {https://doi.org/10.1021/acs.nanolett.1c02392} {\bibfield  {journal}
  {\bibinfo  {journal} {Nano Lett.}\ }\textbf {\bibinfo {volume} {21}},\
  \bibinfo {pages} {6711} (\bibinfo {year} {2021})}\BibitemShut {NoStop}%
\bibitem [{\citenamefont {Wang}\ and\ \citenamefont
  {Sundararaman}(2020)}]{Wang2020}%
  \BibitemOpen
  \bibfield  {author} {\bibinfo {author} {\bibfnamefont {D.}~\bibnamefont
  {Wang}}\ and\ \bibinfo {author} {\bibfnamefont {R.}~\bibnamefont
  {Sundararaman}},\ }\bibfield  {title} {\bibinfo {title} {{Layer dependence of
  defect charge transition levels in two-dimensional materials}},\ }\href
  {https://doi.org/10.1103/physrevb.101.054103} {\bibfield  {journal} {\bibinfo
   {journal} {Phys. Rev. B}\ }\textbf {\bibinfo {volume} {101}},\ \bibinfo
  {pages} {054103} (\bibinfo {year} {2020})}\BibitemShut {NoStop}%
\bibitem [{\citenamefont {Tongay}\ \emph
  {et~al.}(2014{\natexlab{a}})\citenamefont {Tongay}, \citenamefont {Fan},
  \citenamefont {Kang}, \citenamefont {Park}, \citenamefont {Koldemir},
  \citenamefont {Suh}, \citenamefont {Narang}, \citenamefont {Liu},
  \citenamefont {Ji}, \citenamefont {Li}, \citenamefont {Sinclair},\ and\
  \citenamefont {Wu}}]{Tongay2014}%
  \BibitemOpen
  \bibfield  {author} {\bibinfo {author} {\bibfnamefont {S.}~\bibnamefont
  {Tongay}}, \bibinfo {author} {\bibfnamefont {W.}~\bibnamefont {Fan}},
  \bibinfo {author} {\bibfnamefont {J.}~\bibnamefont {Kang}}, \bibinfo {author}
  {\bibfnamefont {J.}~\bibnamefont {Park}}, \bibinfo {author} {\bibfnamefont
  {U.}~\bibnamefont {Koldemir}}, \bibinfo {author} {\bibfnamefont
  {J.}~\bibnamefont {Suh}}, \bibinfo {author} {\bibfnamefont {D.~S.}\
  \bibnamefont {Narang}}, \bibinfo {author} {\bibfnamefont {K.}~\bibnamefont
  {Liu}}, \bibinfo {author} {\bibfnamefont {J.}~\bibnamefont {Ji}}, \bibinfo
  {author} {\bibfnamefont {J.}~\bibnamefont {Li}}, \bibinfo {author}
  {\bibfnamefont {R.}~\bibnamefont {Sinclair}},\ and\ \bibinfo {author}
  {\bibfnamefont {J.}~\bibnamefont {Wu}},\ }\bibfield  {title} {\bibinfo
  {title} {{Tuning Interlayer Coupling in Large-Area Heterostructures with
  CVD-Grown MoS$_2$ and WS$_2$ Monolayers}},\ }\href
  {https://doi.org/10.1021/nl500515q} {\bibfield  {journal} {\bibinfo
  {journal} {Nano Lett.}\ }\textbf {\bibinfo {volume} {14}},\ \bibinfo {pages}
  {3185–3190} (\bibinfo {year} {2014}{\natexlab{a}})}\BibitemShut {NoStop}%
\bibitem [{\citenamefont {Manzeli}\ \emph {et~al.}(2017)\citenamefont
  {Manzeli}, \citenamefont {Ovchinnikov}, \citenamefont {Pasquier},
  \citenamefont {Yazyev},\ and\ \citenamefont {Kis}}]{Manzeli2017}%
  \BibitemOpen
  \bibfield  {author} {\bibinfo {author} {\bibfnamefont {S.}~\bibnamefont
  {Manzeli}}, \bibinfo {author} {\bibfnamefont {D.}~\bibnamefont
  {Ovchinnikov}}, \bibinfo {author} {\bibfnamefont {D.}~\bibnamefont
  {Pasquier}}, \bibinfo {author} {\bibfnamefont {O.~V.}\ \bibnamefont
  {Yazyev}},\ and\ \bibinfo {author} {\bibfnamefont {A.}~\bibnamefont {Kis}},\
  }\bibfield  {title} {\bibinfo {title} {2d transition metal dichalcogenides},\
  }\href {https://doi.org/10.1038/natrevmats.2017.33} {\bibfield  {journal}
  {\bibinfo  {journal} {Nat. Rev. Mater.}\ }\textbf {\bibinfo {volume} {2}},\
  \bibinfo {pages} {17033} (\bibinfo {year} {2017})}\BibitemShut {NoStop}%
\bibitem [{\citenamefont {Meng}\ \emph {et~al.}(2019)\citenamefont {Meng},
  \citenamefont {Pandey}, \citenamefont {Jeong}, \citenamefont {Fu},
  \citenamefont {Yang}, \citenamefont {Chen}, \citenamefont {Singh},
  \citenamefont {He}, \citenamefont {Xu}, \citenamefont {Zhou}, \citenamefont
  {Hsieh}, \citenamefont {Singh}, \citenamefont {Lin},\ and\ \citenamefont
  {Wang}}]{Meng2019}%
  \BibitemOpen
  \bibfield  {author} {\bibinfo {author} {\bibfnamefont {X.}~\bibnamefont
  {Meng}}, \bibinfo {author} {\bibfnamefont {T.}~\bibnamefont {Pandey}},
  \bibinfo {author} {\bibfnamefont {J.}~\bibnamefont {Jeong}}, \bibinfo
  {author} {\bibfnamefont {S.}~\bibnamefont {Fu}}, \bibinfo {author}
  {\bibfnamefont {J.}~\bibnamefont {Yang}}, \bibinfo {author} {\bibfnamefont
  {K.}~\bibnamefont {Chen}}, \bibinfo {author} {\bibfnamefont {A.}~\bibnamefont
  {Singh}}, \bibinfo {author} {\bibfnamefont {F.}~\bibnamefont {He}}, \bibinfo
  {author} {\bibfnamefont {X.}~\bibnamefont {Xu}}, \bibinfo {author}
  {\bibfnamefont {J.}~\bibnamefont {Zhou}}, \bibinfo {author} {\bibfnamefont
  {W.-P.}\ \bibnamefont {Hsieh}}, \bibinfo {author} {\bibfnamefont {A.~K.}\
  \bibnamefont {Singh}}, \bibinfo {author} {\bibfnamefont {J.-F.}\ \bibnamefont
  {Lin}},\ and\ \bibinfo {author} {\bibfnamefont {Y.}~\bibnamefont {Wang}},\
  }\bibfield  {title} {\bibinfo {title} {{Thermal conductivity enhancement in
  MoS$_2$ under extreme strain}},\ }\href
  {https://doi.org/10.1103/physrevlett.122.155901} {\bibfield  {journal}
  {\bibinfo  {journal} {Phys. Rev. Lett.}\ }\textbf {\bibinfo {volume} {122}},\
  \bibinfo {pages} {155901} (\bibinfo {year} {2019})}\BibitemShut {NoStop}%
\bibitem [{\citenamefont {Sun}\ \emph {et~al.}(2017)\citenamefont {Sun},
  \citenamefont {Li}, \citenamefont {Guo}, \citenamefont {Zhao}, \citenamefont
  {Fan}, \citenamefont {Dong}, \citenamefont {Xu}, \citenamefont {Deng},\ and\
  \citenamefont {Fu}}]{Sun2017}%
  \BibitemOpen
  \bibfield  {author} {\bibinfo {author} {\bibfnamefont {J.}~\bibnamefont
  {Sun}}, \bibinfo {author} {\bibfnamefont {X.}~\bibnamefont {Li}}, \bibinfo
  {author} {\bibfnamefont {W.}~\bibnamefont {Guo}}, \bibinfo {author}
  {\bibfnamefont {M.}~\bibnamefont {Zhao}}, \bibinfo {author} {\bibfnamefont
  {X.}~\bibnamefont {Fan}}, \bibinfo {author} {\bibfnamefont {Y.}~\bibnamefont
  {Dong}}, \bibinfo {author} {\bibfnamefont {C.}~\bibnamefont {Xu}}, \bibinfo
  {author} {\bibfnamefont {J.}~\bibnamefont {Deng}},\ and\ \bibinfo {author}
  {\bibfnamefont {Y.}~\bibnamefont {Fu}},\ }\bibfield  {title} {\bibinfo
  {title} {{Synthesis methods of two-dimensional MoS$_2$: a brief review}},\
  }\href {https://doi.org/10.3390/cryst7070198} {\bibfield  {journal} {\bibinfo
   {journal} {Cryst.}\ }\textbf {\bibinfo {volume} {7}},\ \bibinfo {pages}
  {198} (\bibinfo {year} {2017})}\BibitemShut {NoStop}%
\bibitem [{\citenamefont {Collins}\ \emph {et~al.}(1992)\citenamefont
  {Collins}, \citenamefont {Nebesny}, \citenamefont {England}, \citenamefont
  {Chau}, \citenamefont {Lee}, \citenamefont {Parkinson},\ and\ \citenamefont
  {Armstrong}}]{Collins1992}%
  \BibitemOpen
  \bibfield  {author} {\bibinfo {author} {\bibfnamefont {G.~E.}\ \bibnamefont
  {Collins}}, \bibinfo {author} {\bibfnamefont {K.~W.}\ \bibnamefont
  {Nebesny}}, \bibinfo {author} {\bibfnamefont {C.~D.}\ \bibnamefont
  {England}}, \bibinfo {author} {\bibfnamefont {L.-K.}\ \bibnamefont {Chau}},
  \bibinfo {author} {\bibfnamefont {P.~A.}\ \bibnamefont {Lee}}, \bibinfo
  {author} {\bibfnamefont {B.~A.}\ \bibnamefont {Parkinson}},\ and\ \bibinfo
  {author} {\bibfnamefont {N.~R.}\ \bibnamefont {Armstrong}},\ }\bibfield
  {title} {\bibinfo {title} {{Orientation and structure of monolayer -
  multilayer phthalocyanine thin films on layered semiconductor (MoS$_2$ and
  SnS$_2$) surfaces}},\ }\href {https://doi.org/10.1116/1.577728} {\bibfield
  {journal} {\bibinfo  {journal} {J. Vac. Sci. Technolo. A}\ }\textbf {\bibinfo
  {volume} {10}},\ \bibinfo {pages} {2902–2912} (\bibinfo {year}
  {1992})}\BibitemShut {NoStop}%
\bibitem [{\citenamefont {Perea-L{\'o}pez}\ \emph {et~al.}(2014)\citenamefont
  {Perea-L{\'o}pez}, \citenamefont {Lin}, \citenamefont {Pradhan},
  \citenamefont {I{\~n}iguez-R{\'a}bago}, \citenamefont {El{\'\i}as},
  \citenamefont {McCreary}, \citenamefont {Lou}, \citenamefont {Ajayan},
  \citenamefont {Terrones}, \citenamefont {Balicas},\ and\ \citenamefont
  {Terrones}}]{PereaLpez2014}%
  \BibitemOpen
  \bibfield  {author} {\bibinfo {author} {\bibfnamefont {N.}~\bibnamefont
  {Perea-L{\'o}pez}}, \bibinfo {author} {\bibfnamefont {Z.}~\bibnamefont
  {Lin}}, \bibinfo {author} {\bibfnamefont {N.~R.}\ \bibnamefont {Pradhan}},
  \bibinfo {author} {\bibfnamefont {A.}~\bibnamefont {I{\~n}iguez-R{\'a}bago}},
  \bibinfo {author} {\bibfnamefont {A.~L.}\ \bibnamefont {El{\'\i}as}},
  \bibinfo {author} {\bibfnamefont {A.}~\bibnamefont {McCreary}}, \bibinfo
  {author} {\bibfnamefont {J.}~\bibnamefont {Lou}}, \bibinfo {author}
  {\bibfnamefont {P.~M.}\ \bibnamefont {Ajayan}}, \bibinfo {author}
  {\bibfnamefont {H.}~\bibnamefont {Terrones}}, \bibinfo {author}
  {\bibfnamefont {L.}~\bibnamefont {Balicas}},\ and\ \bibinfo {author}
  {\bibfnamefont {M.}~\bibnamefont {Terrones}},\ }\bibfield  {title} {\bibinfo
  {title} {{CVD-grown monolayered MoS$_2$ as an effective photosensor operating
  at low-voltage}},\ }\href {https://doi.org/10.1088/2053-1583/1/1/011004}
  {\bibfield  {journal} {\bibinfo  {journal} {2D Mater.}\ }\textbf {\bibinfo
  {volume} {1}},\ \bibinfo {pages} {011004} (\bibinfo {year}
  {2014})}\BibitemShut {NoStop}%
\bibitem [{\citenamefont {Kim}\ \emph {et~al.}(2022)\citenamefont {Kim},
  \citenamefont {Maity}, \citenamefont {Kim}, \citenamefont {Fu}, \citenamefont
  {Juneja}, \citenamefont {Singh}, \citenamefont {Akinwande},\ and\
  \citenamefont {Lin}}]{Kim2022}%
  \BibitemOpen
  \bibfield  {author} {\bibinfo {author} {\bibfnamefont {J.-S.}\ \bibnamefont
  {Kim}}, \bibinfo {author} {\bibfnamefont {N.}~\bibnamefont {Maity}}, \bibinfo
  {author} {\bibfnamefont {M.}~\bibnamefont {Kim}}, \bibinfo {author}
  {\bibfnamefont {S.}~\bibnamefont {Fu}}, \bibinfo {author} {\bibfnamefont
  {R.}~\bibnamefont {Juneja}}, \bibinfo {author} {\bibfnamefont
  {A.}~\bibnamefont {Singh}}, \bibinfo {author} {\bibfnamefont
  {D.}~\bibnamefont {Akinwande}},\ and\ \bibinfo {author} {\bibfnamefont
  {J.-F.}\ \bibnamefont {Lin}},\ }\bibfield  {title} {\bibinfo {title}
  {{Strain-Modulated Interlayer Charge and Energy Transfers in MoS$_2$/WS$_2$
  Heterobilayer}},\ }\href {https://doi.org/10.1021/acsami.2c10982} {\bibfield
  {journal} {\bibinfo  {journal} {ACS Appl. Mater. Interfaces}\ }\textbf
  {\bibinfo {volume} {14}},\ \bibinfo {pages} {46841–46849} (\bibinfo {year}
  {2022})}\BibitemShut {NoStop}%
\bibitem [{\citenamefont {Yu}\ \emph {et~al.}(2013)\citenamefont {Yu},
  \citenamefont {Li}, \citenamefont {Liu}, \citenamefont {Su}, \citenamefont
  {Zhang},\ and\ \citenamefont {Cao}}]{Yu2013}%
  \BibitemOpen
  \bibfield  {author} {\bibinfo {author} {\bibfnamefont {Y.}~\bibnamefont
  {Yu}}, \bibinfo {author} {\bibfnamefont {C.}~\bibnamefont {Li}}, \bibinfo
  {author} {\bibfnamefont {Y.}~\bibnamefont {Liu}}, \bibinfo {author}
  {\bibfnamefont {L.}~\bibnamefont {Su}}, \bibinfo {author} {\bibfnamefont
  {Y.}~\bibnamefont {Zhang}},\ and\ \bibinfo {author} {\bibfnamefont
  {L.}~\bibnamefont {Cao}},\ }\bibfield  {title} {\bibinfo {title} {{Controlled
  scalable synthesis of uniform, high-quality monolayer and few-layer MoS$_2$
  films}},\ }\href {https://doi.org/10.1038/srep01866} {\bibfield  {journal}
  {\bibinfo  {journal} {Sci. Rep.}\ }\textbf {\bibinfo {volume} {3}},\ \bibinfo
  {pages} {1866} (\bibinfo {year} {2013})}\BibitemShut {NoStop}%
\bibitem [{\citenamefont {Rathi}\ \emph {et~al.}(2015)\citenamefont {Rathi},
  \citenamefont {Lee}, \citenamefont {Lim}, \citenamefont {Wang}, \citenamefont
  {Ochiai}, \citenamefont {Aoki}, \citenamefont {Watanabe}, \citenamefont
  {Taniguchi}, \citenamefont {Lee}, \citenamefont {Yu}, \citenamefont {Kim},\
  and\ \citenamefont {Kim}}]{Rathi2015}%
  \BibitemOpen
  \bibfield  {author} {\bibinfo {author} {\bibfnamefont {S.}~\bibnamefont
  {Rathi}}, \bibinfo {author} {\bibfnamefont {I.}~\bibnamefont {Lee}}, \bibinfo
  {author} {\bibfnamefont {D.}~\bibnamefont {Lim}}, \bibinfo {author}
  {\bibfnamefont {J.}~\bibnamefont {Wang}}, \bibinfo {author} {\bibfnamefont
  {Y.}~\bibnamefont {Ochiai}}, \bibinfo {author} {\bibfnamefont
  {N.}~\bibnamefont {Aoki}}, \bibinfo {author} {\bibfnamefont {K.}~\bibnamefont
  {Watanabe}}, \bibinfo {author} {\bibfnamefont {T.}~\bibnamefont {Taniguchi}},
  \bibinfo {author} {\bibfnamefont {G.-H.}\ \bibnamefont {Lee}}, \bibinfo
  {author} {\bibfnamefont {Y.-J.}\ \bibnamefont {Yu}}, \bibinfo {author}
  {\bibfnamefont {P.}~\bibnamefont {Kim}},\ and\ \bibinfo {author}
  {\bibfnamefont {G.-H.}\ \bibnamefont {Kim}},\ }\bibfield  {title} {\bibinfo
  {title} {{Tunable electrical and optical characteristics in monolayer
  graphene and few-layer MoS$_2$ heterostructure devices}},\ }\href
  {https://doi.org/10.1021/acs.nanolett.5b01030} {\bibfield  {journal}
  {\bibinfo  {journal} {Nano Lett.}\ }\textbf {\bibinfo {volume} {15}},\
  \bibinfo {pages} {5017–5024} (\bibinfo {year} {2015})}\BibitemShut
  {NoStop}%
\bibitem [{\citenamefont {Yu}\ \emph {et~al.}(2014)\citenamefont {Yu},
  \citenamefont {Huang}, \citenamefont {Li}, \citenamefont {Steinmann},
  \citenamefont {Yang},\ and\ \citenamefont {Cao}}]{Yu2014}%
  \BibitemOpen
  \bibfield  {author} {\bibinfo {author} {\bibfnamefont {Y.}~\bibnamefont
  {Yu}}, \bibinfo {author} {\bibfnamefont {S.-Y.}\ \bibnamefont {Huang}},
  \bibinfo {author} {\bibfnamefont {Y.}~\bibnamefont {Li}}, \bibinfo {author}
  {\bibfnamefont {S.~N.}\ \bibnamefont {Steinmann}}, \bibinfo {author}
  {\bibfnamefont {W.}~\bibnamefont {Yang}},\ and\ \bibinfo {author}
  {\bibfnamefont {L.}~\bibnamefont {Cao}},\ }\bibfield  {title} {\bibinfo
  {title} {{Layer-dependent electrocatalysis of MoS$_2$ for hydrogen
  evolution}},\ }\href {https://doi.org/10.1021/nl403620g} {\bibfield
  {journal} {\bibinfo  {journal} {Nano Lett.}\ }\textbf {\bibinfo {volume}
  {14}},\ \bibinfo {pages} {553–558} (\bibinfo {year} {2014})}\BibitemShut
  {NoStop}%
\bibitem [{\citenamefont {Li}\ \emph {et~al.}(2017)\citenamefont {Li},
  \citenamefont {Han}, \citenamefont {Wu}, \citenamefont {Qiao}, \citenamefont
  {Zhang},\ and\ \citenamefont {Tan}}]{Li2017}%
  \BibitemOpen
  \bibfield  {author} {\bibinfo {author} {\bibfnamefont {X.}~\bibnamefont
  {Li}}, \bibinfo {author} {\bibfnamefont {W.}~\bibnamefont {Han}}, \bibinfo
  {author} {\bibfnamefont {J.}~\bibnamefont {Wu}}, \bibinfo {author}
  {\bibfnamefont {X.}~\bibnamefont {Qiao}}, \bibinfo {author} {\bibfnamefont
  {J.}~\bibnamefont {Zhang}},\ and\ \bibinfo {author} {\bibfnamefont
  {P.}~\bibnamefont {Tan}},\ }\bibfield  {title} {\bibinfo {title}
  {{Layer-number dependent optical properties of 2D materials and their
  application for thickness determination}},\ }\href
  {https://doi.org/10.1002/adfm.201604468} {\bibfield  {journal} {\bibinfo
  {journal} {Adv. Funct. Mater.}\ }\textbf {\bibinfo {volume} {27}},\ \bibinfo
  {pages} {1604468} (\bibinfo {year} {2017})}\BibitemShut {NoStop}%
\bibitem [{\citenamefont {Hong}\ \emph {et~al.}(2016)\citenamefont {Hong},
  \citenamefont {Li}, \citenamefont {Jin}, \citenamefont {Zhang}, \citenamefont
  {Zhang},\ and\ \citenamefont {Yuan}}]{Hong2016}%
  \BibitemOpen
  \bibfield  {author} {\bibinfo {author} {\bibfnamefont {J.}~\bibnamefont
  {Hong}}, \bibinfo {author} {\bibfnamefont {K.}~\bibnamefont {Li}}, \bibinfo
  {author} {\bibfnamefont {C.}~\bibnamefont {Jin}}, \bibinfo {author}
  {\bibfnamefont {X.}~\bibnamefont {Zhang}}, \bibinfo {author} {\bibfnamefont
  {Z.}~\bibnamefont {Zhang}},\ and\ \bibinfo {author} {\bibfnamefont
  {J.}~\bibnamefont {Yuan}},\ }\bibfield  {title} {\bibinfo {title}
  {{Layer-dependent anisotropic electronic structure of freestanding
  quasi-two-dimensional MoS$_2$}},\ }\href
  {https://doi.org/10.1103/physrevb.93.075440} {\bibfield  {journal} {\bibinfo
  {journal} {Phys. Rev. B}\ }\textbf {\bibinfo {volume} {93}},\ \bibinfo
  {pages} {075440} (\bibinfo {year} {2016})}\BibitemShut {NoStop}%
\bibitem [{\citenamefont {Tongay}\ \emph
  {et~al.}(2014{\natexlab{b}})\citenamefont {Tongay}, \citenamefont {Sahin},
  \citenamefont {Ko}, \citenamefont {Luce}, \citenamefont {Fan}, \citenamefont
  {Liu}, \citenamefont {Zhou}, \citenamefont {Huang}, \citenamefont {Ho},
  \citenamefont {Yan}, \citenamefont {Ogletree}, \citenamefont {Aloni},
  \citenamefont {Ji}, \citenamefont {Li}, \citenamefont {Li}, \citenamefont
  {Peeters},\ and\ \citenamefont {Wu}}]{Tongay2014natcomm}%
  \BibitemOpen
  \bibfield  {author} {\bibinfo {author} {\bibfnamefont {S.}~\bibnamefont
  {Tongay}}, \bibinfo {author} {\bibfnamefont {H.}~\bibnamefont {Sahin}},
  \bibinfo {author} {\bibfnamefont {C.}~\bibnamefont {Ko}}, \bibinfo {author}
  {\bibfnamefont {A.}~\bibnamefont {Luce}}, \bibinfo {author} {\bibfnamefont
  {W.}~\bibnamefont {Fan}}, \bibinfo {author} {\bibfnamefont {K.}~\bibnamefont
  {Liu}}, \bibinfo {author} {\bibfnamefont {J.}~\bibnamefont {Zhou}}, \bibinfo
  {author} {\bibfnamefont {Y.-S.}\ \bibnamefont {Huang}}, \bibinfo {author}
  {\bibfnamefont {C.-H.}\ \bibnamefont {Ho}}, \bibinfo {author} {\bibfnamefont
  {J.}~\bibnamefont {Yan}}, \bibinfo {author} {\bibfnamefont {D.~F.}\
  \bibnamefont {Ogletree}}, \bibinfo {author} {\bibfnamefont {S.}~\bibnamefont
  {Aloni}}, \bibinfo {author} {\bibfnamefont {J.}~\bibnamefont {Ji}}, \bibinfo
  {author} {\bibfnamefont {S.}~\bibnamefont {Li}}, \bibinfo {author}
  {\bibfnamefont {J.}~\bibnamefont {Li}}, \bibinfo {author} {\bibfnamefont
  {F.~M.}\ \bibnamefont {Peeters}},\ and\ \bibinfo {author} {\bibfnamefont
  {J.}~\bibnamefont {Wu}},\ }\bibfield  {title} {\bibinfo {title} {{Monolayer
  behaviour in bulk ReS$_2$ due to electronic and vibrational decoupling}},\
  }\href {https://doi.org/10.1038/ncomms4252} {\bibfield  {journal} {\bibinfo
  {journal} {Nat. Commun.}\ }\textbf {\bibinfo {volume} {5}},\ \bibinfo {pages}
  {1} (\bibinfo {year} {2014}{\natexlab{b}})}\BibitemShut {NoStop}%
\bibitem [{\citenamefont {Chenet}\ \emph {et~al.}(2015)\citenamefont {Chenet},
  \citenamefont {Aslan}, \citenamefont {Huang}, \citenamefont {Fan},
  \citenamefont {van~der Zande}, \citenamefont {Heinz},\ and\ \citenamefont
  {Hone}}]{Chenet2015}%
  \BibitemOpen
  \bibfield  {author} {\bibinfo {author} {\bibfnamefont {D.~A.}\ \bibnamefont
  {Chenet}}, \bibinfo {author} {\bibfnamefont {B.}~\bibnamefont {Aslan}},
  \bibinfo {author} {\bibfnamefont {P.~Y.}\ \bibnamefont {Huang}}, \bibinfo
  {author} {\bibfnamefont {C.}~\bibnamefont {Fan}}, \bibinfo {author}
  {\bibfnamefont {A.~M.}\ \bibnamefont {van~der Zande}}, \bibinfo {author}
  {\bibfnamefont {T.~F.}\ \bibnamefont {Heinz}},\ and\ \bibinfo {author}
  {\bibfnamefont {J.~C.}\ \bibnamefont {Hone}},\ }\bibfield  {title} {\bibinfo
  {title} {{In-Plane Anisotropy in Mono- and Few-Layer ReS$_2$ Probed by Raman
  Spectroscopy and Scanning Transmission Electron Microscopy}},\ }\href
  {https://doi.org/10.1021/acs.nanolett.5b00910} {\bibfield  {journal}
  {\bibinfo  {journal} {Nano Lett.}\ }\textbf {\bibinfo {volume} {15}},\
  \bibinfo {pages} {5667–5672} (\bibinfo {year} {2015})}\BibitemShut
  {NoStop}%
\bibitem [{\citenamefont {Jariwala}\ \emph {et~al.}(2016)\citenamefont
  {Jariwala}, \citenamefont {Voiry}, \citenamefont {Jindal}, \citenamefont
  {Chalke}, \citenamefont {Bapat}, \citenamefont {Thamizhavel}, \citenamefont
  {Chhowalla}, \citenamefont {Deshmukh},\ and\ \citenamefont
  {Bhattacharya}}]{Jariwala2016}%
  \BibitemOpen
  \bibfield  {author} {\bibinfo {author} {\bibfnamefont {B.}~\bibnamefont
  {Jariwala}}, \bibinfo {author} {\bibfnamefont {D.}~\bibnamefont {Voiry}},
  \bibinfo {author} {\bibfnamefont {A.}~\bibnamefont {Jindal}}, \bibinfo
  {author} {\bibfnamefont {B.~A.}\ \bibnamefont {Chalke}}, \bibinfo {author}
  {\bibfnamefont {R.}~\bibnamefont {Bapat}}, \bibinfo {author} {\bibfnamefont
  {A.}~\bibnamefont {Thamizhavel}}, \bibinfo {author} {\bibfnamefont
  {M.}~\bibnamefont {Chhowalla}}, \bibinfo {author} {\bibfnamefont
  {M.}~\bibnamefont {Deshmukh}},\ and\ \bibinfo {author} {\bibfnamefont
  {A.}~\bibnamefont {Bhattacharya}},\ }\bibfield  {title} {\bibinfo {title}
  {{Synthesis and characterization of ReS$_2$ and ReSe$_2$ layered chalcogenide
  single crystals}},\ }\href {https://doi.org/10.1021/acs.chemmater.6b00364}
  {\bibfield  {journal} {\bibinfo  {journal} {Chem. Mater.}\ }\textbf {\bibinfo
  {volume} {28}},\ \bibinfo {pages} {3352} (\bibinfo {year}
  {2016})}\BibitemShut {NoStop}%
\bibitem [{\citenamefont {Feng}\ \emph {et~al.}(2015)\citenamefont {Feng},
  \citenamefont {Zhou}, \citenamefont {Wang}, \citenamefont {Zhou},
  \citenamefont {Liu}, \citenamefont {Fu}, \citenamefont {Ni}, \citenamefont
  {Wu}, \citenamefont {Yuan}, \citenamefont {Miao}, \citenamefont {Wang},
  \citenamefont {Wan},\ and\ \citenamefont {Xing}}]{Feng2015}%
  \BibitemOpen
  \bibfield  {author} {\bibinfo {author} {\bibfnamefont {Y.}~\bibnamefont
  {Feng}}, \bibinfo {author} {\bibfnamefont {W.}~\bibnamefont {Zhou}}, \bibinfo
  {author} {\bibfnamefont {Y.}~\bibnamefont {Wang}}, \bibinfo {author}
  {\bibfnamefont {J.}~\bibnamefont {Zhou}}, \bibinfo {author} {\bibfnamefont
  {E.}~\bibnamefont {Liu}}, \bibinfo {author} {\bibfnamefont {Y.}~\bibnamefont
  {Fu}}, \bibinfo {author} {\bibfnamefont {Z.}~\bibnamefont {Ni}}, \bibinfo
  {author} {\bibfnamefont {X.}~\bibnamefont {Wu}}, \bibinfo {author}
  {\bibfnamefont {H.}~\bibnamefont {Yuan}}, \bibinfo {author} {\bibfnamefont
  {F.}~\bibnamefont {Miao}}, \bibinfo {author} {\bibfnamefont {B.}~\bibnamefont
  {Wang}}, \bibinfo {author} {\bibfnamefont {X.}~\bibnamefont {Wan}},\ and\
  \bibinfo {author} {\bibfnamefont {D.}~\bibnamefont {Xing}},\ }\bibfield
  {title} {\bibinfo {title} {{Raman vibrational spectra of bulk to monolayer
  ReS$_2$ with lower symmetry}},\ }\href
  {https://doi.org/10.1103/physrevb.92.054110} {\bibfield  {journal} {\bibinfo
  {journal} {Phys. Rev. B}\ }\textbf {\bibinfo {volume} {92}},\ \bibinfo
  {pages} {054110} (\bibinfo {year} {2015})}\BibitemShut {NoStop}%
\bibitem [{\citenamefont {Zhou}\ \emph {et~al.}(2020)\citenamefont {Zhou},
  \citenamefont {Maity}, \citenamefont {Rai}, \citenamefont {Juneja},
  \citenamefont {Meng}, \citenamefont {Roy}, \citenamefont {Zhang},
  \citenamefont {Xu}, \citenamefont {Lin}, \citenamefont {Banerjee},
  \citenamefont {Singh},\ and\ \citenamefont {Wang}}]{Zhou2020}%
  \BibitemOpen
  \bibfield  {author} {\bibinfo {author} {\bibfnamefont {Y.}~\bibnamefont
  {Zhou}}, \bibinfo {author} {\bibfnamefont {N.}~\bibnamefont {Maity}},
  \bibinfo {author} {\bibfnamefont {A.}~\bibnamefont {Rai}}, \bibinfo {author}
  {\bibfnamefont {R.}~\bibnamefont {Juneja}}, \bibinfo {author} {\bibfnamefont
  {X.}~\bibnamefont {Meng}}, \bibinfo {author} {\bibfnamefont {A.}~\bibnamefont
  {Roy}}, \bibinfo {author} {\bibfnamefont {Y.}~\bibnamefont {Zhang}}, \bibinfo
  {author} {\bibfnamefont {X.}~\bibnamefont {Xu}}, \bibinfo {author}
  {\bibfnamefont {J.}~\bibnamefont {Lin}}, \bibinfo {author} {\bibfnamefont
  {S.~K.}\ \bibnamefont {Banerjee}}, \bibinfo {author} {\bibfnamefont {A.~K.}\
  \bibnamefont {Singh}},\ and\ \bibinfo {author} {\bibfnamefont
  {Y.}~\bibnamefont {Wang}},\ }\bibfield  {title} {\bibinfo {title}
  {{Stacking-Order-Driven Optical Properties and Carrier Dynamics in
  ReS$_2$}},\ }\href {https://doi.org/10.1002/adma.201908311} {\bibfield
  {journal} {\bibinfo  {journal} {Adv. Mater.}\ }\textbf {\bibinfo {volume}
  {32}},\ \bibinfo {pages} {1908311} (\bibinfo {year} {2020})}\BibitemShut
  {NoStop}%
\bibitem [{\citenamefont {Kresse}\ and\ \citenamefont
  {Furthm\"{u}ller}(1996{\natexlab{a}})}]{Kresse1996}%
  \BibitemOpen
  \bibfield  {author} {\bibinfo {author} {\bibfnamefont {G.}~\bibnamefont
  {Kresse}}\ and\ \bibinfo {author} {\bibfnamefont {J.}~\bibnamefont
  {Furthm\"{u}ller}},\ }\bibfield  {title} {\bibinfo {title} {{Efficient
  iterative schemes for ab initio total-energy calculations using a plane-wave
  basis set}},\ }\href {https://doi.org/10.1103/physrevb.54.11169} {\bibfield
  {journal} {\bibinfo  {journal} {Phys. Rev. B}\ }\textbf {\bibinfo {volume}
  {54}},\ \bibinfo {pages} {11169} (\bibinfo {year}
  {1996}{\natexlab{a}})}\BibitemShut {NoStop}%
\bibitem [{\citenamefont {Kresse}\ and\ \citenamefont
  {Furthm\"{u}ller}(1996{\natexlab{b}})}]{Kresse1996comp}%
  \BibitemOpen
  \bibfield  {author} {\bibinfo {author} {\bibfnamefont {G.}~\bibnamefont
  {Kresse}}\ and\ \bibinfo {author} {\bibfnamefont {J.}~\bibnamefont
  {Furthm\"{u}ller}},\ }\bibfield  {title} {\bibinfo {title} {{Efficiency of
  ab-initio total energy calculations for metals and semiconductors using a
  plane-wave basis set}},\ }\href
  {https://doi.org/10.1016/0927-0256(96)00008-0} {\bibfield  {journal}
  {\bibinfo  {journal} {Comput. Mater. Sci.}\ }\textbf {\bibinfo {volume}
  {6}},\ \bibinfo {pages} {15} (\bibinfo {year}
  {1996}{\natexlab{b}})}\BibitemShut {NoStop}%
\bibitem [{\citenamefont {Bl\"{o}chl}(1994)}]{Blochl1994}%
  \BibitemOpen
  \bibfield  {author} {\bibinfo {author} {\bibfnamefont {P.~E.}\ \bibnamefont
  {Bl\"{o}chl}},\ }\bibfield  {title} {\bibinfo {title} {Projector
  augmented-wave method},\ }\href {https://doi.org/10.1103/physrevb.50.17953}
  {\bibfield  {journal} {\bibinfo  {journal} {Phys. Rev. B}\ }\textbf {\bibinfo
  {volume} {50}},\ \bibinfo {pages} {17953–17979} (\bibinfo {year}
  {1994})}\BibitemShut {NoStop}%
\bibitem [{\citenamefont {Kresse}\ and\ \citenamefont
  {Joubert}(1999)}]{Kresse1999}%
  \BibitemOpen
  \bibfield  {author} {\bibinfo {author} {\bibfnamefont {G.}~\bibnamefont
  {Kresse}}\ and\ \bibinfo {author} {\bibfnamefont {D.}~\bibnamefont
  {Joubert}},\ }\bibfield  {title} {\bibinfo {title} {From ultrasoft
  pseudopotentials to the projector augmented-wave method},\ }\href
  {https://doi.org/10.1103/physrevb.59.1758} {\bibfield  {journal} {\bibinfo
  {journal} {Phys. Rev. B}\ }\textbf {\bibinfo {volume} {59}},\ \bibinfo
  {pages} {1758} (\bibinfo {year} {1999})}\BibitemShut {NoStop}%
\bibitem [{\citenamefont {Perdew}\ \emph {et~al.}(1996)\citenamefont {Perdew},
  \citenamefont {Burke},\ and\ \citenamefont {Ernzerhof}}]{Perdew1996}%
  \BibitemOpen
  \bibfield  {author} {\bibinfo {author} {\bibfnamefont {J.~P.}\ \bibnamefont
  {Perdew}}, \bibinfo {author} {\bibfnamefont {K.}~\bibnamefont {Burke}},\ and\
  \bibinfo {author} {\bibfnamefont {M.}~\bibnamefont {Ernzerhof}},\ }\bibfield
  {title} {\bibinfo {title} {{Generalized Gradient Approximation Made
  Simple}},\ }\href {https://doi.org/10.1103/physrevlett.77.3865} {\bibfield
  {journal} {\bibinfo  {journal} {Phys. Rev. Lett.}\ }\textbf {\bibinfo
  {volume} {77}},\ \bibinfo {pages} {3865–3868} (\bibinfo {year}
  {1996})}\BibitemShut {NoStop}%
\bibitem [{\citenamefont {Klimeš}\ \emph {et~al.}(2009)\citenamefont
  {Klimeš}, \citenamefont {Bowler},\ and\ \citenamefont
  {Michaelides}}]{Klime2009}%
  \BibitemOpen
  \bibfield  {author} {\bibinfo {author} {\bibfnamefont {J.}~\bibnamefont
  {Klimeš}}, \bibinfo {author} {\bibfnamefont {D.~R.}\ \bibnamefont
  {Bowler}},\ and\ \bibinfo {author} {\bibfnamefont {A.}~\bibnamefont
  {Michaelides}},\ }\bibfield  {title} {\bibinfo {title} {Chemical accuracy for
  the van der waals density functional},\ }\href
  {https://doi.org/10.1088/0953-8984/22/2/022201} {\bibfield  {journal}
  {\bibinfo  {journal} {J. Phys. Condens. Matter}\ }\textbf {\bibinfo {volume}
  {22}},\ \bibinfo {pages} {022201} (\bibinfo {year} {2009})}\BibitemShut
  {NoStop}%
\bibitem [{\citenamefont {Klimeš}\ \emph {et~al.}(2011)\citenamefont
  {Klimeš}, \citenamefont {Bowler},\ and\ \citenamefont
  {Michaelides}}]{Klime2011}%
  \BibitemOpen
  \bibfield  {author} {\bibinfo {author} {\bibfnamefont {J.}~\bibnamefont
  {Klimeš}}, \bibinfo {author} {\bibfnamefont {D.~R.}\ \bibnamefont
  {Bowler}},\ and\ \bibinfo {author} {\bibfnamefont {A.}~\bibnamefont
  {Michaelides}},\ }\bibfield  {title} {\bibinfo {title} {Van der waals density
  functionals applied to solids},\ }\href
  {https://doi.org/10.1103/physrevb.83.195131} {\bibfield  {journal} {\bibinfo
  {journal} {Phys. Rev. B}\ }\textbf {\bibinfo {volume} {83}},\ \bibinfo
  {pages} {195131} (\bibinfo {year} {2011})}\BibitemShut {NoStop}%
\bibitem [{\citenamefont {Monkhorst}\ and\ \citenamefont
  {Pack}(1976)}]{Monkhorst1976}%
  \BibitemOpen
  \bibfield  {author} {\bibinfo {author} {\bibfnamefont {H.~J.}\ \bibnamefont
  {Monkhorst}}\ and\ \bibinfo {author} {\bibfnamefont {J.~D.}\ \bibnamefont
  {Pack}},\ }\bibfield  {title} {\bibinfo {title} {{Special points for
  Brillouin-zone integrations}},\ }\href
  {https://doi.org/10.1103/physrevb.13.5188} {\bibfield  {journal} {\bibinfo
  {journal} {Phys. Rev. B}\ }\textbf {\bibinfo {volume} {13}},\ \bibinfo
  {pages} {5188} (\bibinfo {year} {1976})}\BibitemShut {NoStop}%
\bibitem [{\citenamefont {Freysoldt}\ \emph {et~al.}(2014)\citenamefont
  {Freysoldt}, \citenamefont {Grabowski}, \citenamefont {Hickel}, \citenamefont
  {Neugebauer}, \citenamefont {Kresse}, \citenamefont {Janotti},\ and\
  \citenamefont {Van~de Walle}}]{Freysoldt2014}%
  \BibitemOpen
  \bibfield  {author} {\bibinfo {author} {\bibfnamefont {C.}~\bibnamefont
  {Freysoldt}}, \bibinfo {author} {\bibfnamefont {B.}~\bibnamefont
  {Grabowski}}, \bibinfo {author} {\bibfnamefont {T.}~\bibnamefont {Hickel}},
  \bibinfo {author} {\bibfnamefont {J.}~\bibnamefont {Neugebauer}}, \bibinfo
  {author} {\bibfnamefont {G.}~\bibnamefont {Kresse}}, \bibinfo {author}
  {\bibfnamefont {A.}~\bibnamefont {Janotti}},\ and\ \bibinfo {author}
  {\bibfnamefont {C.~G.}\ \bibnamefont {Van~de Walle}},\ }\bibfield  {title}
  {\bibinfo {title} {First-principles calculations for point defects in
  solids},\ }\href {https://doi.org/10.1103/revmodphys.86.253} {\bibfield
  {journal} {\bibinfo  {journal} {Rev. Mod. Phys.}\ }\textbf {\bibinfo {volume}
  {86}},\ \bibinfo {pages} {253} (\bibinfo {year} {2014})}\BibitemShut
  {NoStop}%
\bibitem [{\citenamefont {Van~de Walle}\ and\ \citenamefont
  {Neugebauer}(2004)}]{VandeWalle2004}%
  \BibitemOpen
  \bibfield  {author} {\bibinfo {author} {\bibfnamefont {C.~G.}\ \bibnamefont
  {Van~de Walle}}\ and\ \bibinfo {author} {\bibfnamefont {J.}~\bibnamefont
  {Neugebauer}},\ }\bibfield  {title} {\bibinfo {title} {{First-principles
  calculations for defects and impurities: Applications to III-nitrides}},\
  }\href {https://doi.org/10.1063/1.1682673} {\bibfield  {journal} {\bibinfo
  {journal} {J. Appl. Phys.}\ }\textbf {\bibinfo {volume} {95}},\ \bibinfo
  {pages} {3851} (\bibinfo {year} {2004})}\BibitemShut {NoStop}%
\bibitem [{\citenamefont {Dey}\ \emph {et~al.}(2021)\citenamefont {Dey},
  \citenamefont {Singh},\ and\ \citenamefont {Singh}}]{Dey2021}%
  \BibitemOpen
  \bibfield  {author} {\bibinfo {author} {\bibfnamefont {M.}~\bibnamefont
  {Dey}}, \bibinfo {author} {\bibfnamefont {A.}~\bibnamefont {Singh}},\ and\
  \bibinfo {author} {\bibfnamefont {A.~K.}\ \bibnamefont {Singh}},\ }\bibfield
  {title} {\bibinfo {title} {Formation of a small electron polaron in tantalum
  oxynitride: Origin of low mobility},\ }\href
  {https://doi.org/10.1021/acs.jpcc.1c00702} {\bibfield  {journal} {\bibinfo
  {journal} {J. Phys. Chem. C}\ }\textbf {\bibinfo {volume} {125}},\ \bibinfo
  {pages} {11548} (\bibinfo {year} {2021})}\BibitemShut {NoStop}%
\bibitem [{\citenamefont {Dey}\ and\ \citenamefont {Singh}(2023)}]{Dey2023}%
  \BibitemOpen
  \bibfield  {author} {\bibinfo {author} {\bibfnamefont {M.}~\bibnamefont
  {Dey}}\ and\ \bibinfo {author} {\bibfnamefont {A.~K.}\ \bibnamefont
  {Singh}},\ }\bibfield  {title} {\bibinfo {title} {{Broad photoluminescence
  from large Frank-Condon relaxation dynamics of hole polarons in LiGaO$_2$}},\
  }\href {https://doi.org/10.1103/physrevb.108.l041201} {\bibfield  {journal}
  {\bibinfo  {journal} {Phys. Rev. B}\ }\textbf {\bibinfo {volume} {108}},\
  \bibinfo {pages} {L041201} (\bibinfo {year} {2023})}\BibitemShut {NoStop}%
\bibitem [{\citenamefont {Freysoldt}\ \emph {et~al.}(2009)\citenamefont
  {Freysoldt}, \citenamefont {Neugebauer},\ and\ \citenamefont {Van~de
  Walle}}]{Freysoldt2009}%
  \BibitemOpen
  \bibfield  {author} {\bibinfo {author} {\bibfnamefont {C.}~\bibnamefont
  {Freysoldt}}, \bibinfo {author} {\bibfnamefont {J.}~\bibnamefont
  {Neugebauer}},\ and\ \bibinfo {author} {\bibfnamefont {C.~G.}\ \bibnamefont
  {Van~de Walle}},\ }\bibfield  {title} {\bibinfo {title} {{Fully Ab Initio
  Finite-Size Corrections for Charged-Defect Supercell Calculations}},\ }\href
  {https://doi.org/10.1103/physrevlett.102.016402} {\bibfield  {journal}
  {\bibinfo  {journal} {Phys. Rev. Lett.}\ }\textbf {\bibinfo {volume} {102}},\
  \bibinfo {pages} {016402} (\bibinfo {year} {2009})}\BibitemShut {NoStop}%
\bibitem [{\citenamefont {Freysoldt}\ \emph {et~al.}(2010)\citenamefont
  {Freysoldt}, \citenamefont {Neugebauer},\ and\ \citenamefont {Van~de
  Walle}}]{Freysoldt2010}%
  \BibitemOpen
  \bibfield  {author} {\bibinfo {author} {\bibfnamefont {C.}~\bibnamefont
  {Freysoldt}}, \bibinfo {author} {\bibfnamefont {J.}~\bibnamefont
  {Neugebauer}},\ and\ \bibinfo {author} {\bibfnamefont {C.~G.}\ \bibnamefont
  {Van~de Walle}},\ }\bibfield  {title} {\bibinfo {title} {Electrostatic
  interactions between charged defects in supercells},\ }\href
  {https://doi.org/10.1002/pssb.201046289} {\bibfield  {journal} {\bibinfo
  {journal} {physica status solidi (b)}\ }\textbf {\bibinfo {volume} {248}},\
  \bibinfo {pages} {1067} (\bibinfo {year} {2010})}\BibitemShut {NoStop}%
\bibitem [{\citenamefont {Freysoldt}\ and\ \citenamefont
  {Neugebauer}(2018)}]{Freysoldt2018}%
  \BibitemOpen
  \bibfield  {author} {\bibinfo {author} {\bibfnamefont {C.}~\bibnamefont
  {Freysoldt}}\ and\ \bibinfo {author} {\bibfnamefont {J.}~\bibnamefont
  {Neugebauer}},\ }\bibfield  {title} {\bibinfo {title} {First-principles
  calculations for charged defects at surfaces, interfaces, and two-dimensional
  materials in the presence of electric fields},\ }\href
  {https://doi.org/10.1103/physrevb.97.205425} {\bibfield  {journal} {\bibinfo
  {journal} {Phys. Rev. B}\ }\textbf {\bibinfo {volume} {97}},\ \bibinfo
  {pages} {205425} (\bibinfo {year} {2018})}\BibitemShut {NoStop}%
\bibitem [{\citenamefont {Zhou}\ \emph {et~al.}(2021)\citenamefont {Zhou},
  \citenamefont {Maity}, \citenamefont {Lin}, \citenamefont {Singh},\ and\
  \citenamefont {Wang}}]{Zhou2021}%
  \BibitemOpen
  \bibfield  {author} {\bibinfo {author} {\bibfnamefont {Y.}~\bibnamefont
  {Zhou}}, \bibinfo {author} {\bibfnamefont {N.}~\bibnamefont {Maity}},
  \bibinfo {author} {\bibfnamefont {J.-F.}\ \bibnamefont {Lin}}, \bibinfo
  {author} {\bibfnamefont {A.~K.}\ \bibnamefont {Singh}},\ and\ \bibinfo
  {author} {\bibfnamefont {Y.}~\bibnamefont {Wang}},\ }\bibfield  {title}
  {\bibinfo {title} {{Nonlinear Optical Absorption of ReS$_2$ Driven by
  Stacking Order}},\ }\href {https://doi.org/10.1021/acsphotonics.0c01225}
  {\bibfield  {journal} {\bibinfo  {journal} {ACS Photonics}\ }\textbf
  {\bibinfo {volume} {8}},\ \bibinfo {pages} {405} (\bibinfo {year}
  {2021})}\BibitemShut {NoStop}%
\bibitem [{\citenamefont {Upadhyay}\ \emph {et~al.}(2022)\citenamefont
  {Upadhyay}, \citenamefont {Maity}, \citenamefont {Kumar}, \citenamefont
  {Barman}, \citenamefont {Singh},\ and\ \citenamefont {Nayak}}]{Upadhyay2022}%
  \BibitemOpen
  \bibfield  {author} {\bibinfo {author} {\bibfnamefont {P.}~\bibnamefont
  {Upadhyay}}, \bibinfo {author} {\bibfnamefont {N.}~\bibnamefont {Maity}},
  \bibinfo {author} {\bibfnamefont {R.}~\bibnamefont {Kumar}}, \bibinfo
  {author} {\bibfnamefont {P.~K.}\ \bibnamefont {Barman}}, \bibinfo {author}
  {\bibfnamefont {A.~K.}\ \bibnamefont {Singh}},\ and\ \bibinfo {author}
  {\bibfnamefont {P.~K.}\ \bibnamefont {Nayak}},\ }\bibfield  {title} {\bibinfo
  {title} {{Layer parity dependent Raman-active modes and crystal symmetry in
  ReS$_2$}},\ }\href {https://doi.org/10.1103/physrevb.105.045416} {\bibfield
  {journal} {\bibinfo  {journal} {Phys. Rev. B}\ }\textbf {\bibinfo {volume}
  {105}},\ \bibinfo {pages} {045416} (\bibinfo {year} {2022})}\BibitemShut
  {NoStop}%
\bibitem [{SM()}]{SM}%
  \BibitemOpen
  \href@noop {} {\bibinfo  {journal} {See Supplemental Material for additional
  figures and analyses, including: band-edge positions with respect to the
  vacuum level for ReS$_2$ in AA and AB stacking from monolayer to bulk, and
  calculated versus experimental band-gap comparisons; atomic geometries of
  intrinsic point defects (V$_{S1}$, V$_{S2}$, V$_{Re}$, Re$_{S1}$, Re$_{S2}$,
  and S$_{Re}$); schematics of defective-layer placement in multilayer ReS$_2$
  (1L--4L and bulk); formation energies of S$_{Re}$ from monolayer to bulk
  under Re-rich conditions for AA and AB stacking; ionization energies and
  defect transition levels of S$_{Re}$ as functions of layer number; static
  dielectric constants and their correlation with defect transition levels for
  AB stacking; Kohn--Sham levels of all considered defects and layer-dependent
  Kohn--Sham states of S$_{Re}$; electronic and structural relaxation energies
  for AB stacking; and a comparison of interlayer coupling strengths in ReS$_2$
  and ReSe$_2$. The Supplemental Material also contains Refs. [38, 40, 61]}\
  }\BibitemShut {NoStop}%
\bibitem [{\citenamefont {Dileep}\ \emph {et~al.}(2016)\citenamefont {Dileep},
  \citenamefont {Sahu}, \citenamefont {Sarkar}, \citenamefont {Peter},\ and\
  \citenamefont {Datta}}]{Dileep2016}%
  \BibitemOpen
\bibfield  {journal} {  }\bibfield  {author} {\bibinfo {author} {\bibfnamefont
  {K.}~\bibnamefont {Dileep}}, \bibinfo {author} {\bibfnamefont
  {R.}~\bibnamefont {Sahu}}, \bibinfo {author} {\bibfnamefont {S.}~\bibnamefont
  {Sarkar}}, \bibinfo {author} {\bibfnamefont {S.~C.}\ \bibnamefont {Peter}},\
  and\ \bibinfo {author} {\bibfnamefont {R.}~\bibnamefont {Datta}},\ }\bibfield
   {title} {\bibinfo {title} {{Layer specific optical band gap measurement at
  nanoscale in MoS$_2$ and ReS$_2$ van der Waals compounds by high resolution
  electron energy loss spectroscopy}},\ }\href
  {https://doi.org/10.1063/1.4944431} {\bibfield  {journal} {\bibinfo
  {journal} {J. Appl. Phys.}\ }\textbf {\bibinfo {volume} {119}},\ \bibinfo
  {pages} {114309} (\bibinfo {year} {2016})}\BibitemShut {NoStop}%
\bibitem [{\citenamefont {Lin}\ \emph {et~al.}(2015)\citenamefont {Lin},
  \citenamefont {Komsa}, \citenamefont {Yeh}, \citenamefont {Bj\"{o}rkman},
  \citenamefont {Liang}, \citenamefont {Ho}, \citenamefont {Huang},
  \citenamefont {Chiu}, \citenamefont {Krasheninnikov},\ and\ \citenamefont
  {Suenaga}}]{Lin2015}%
  \BibitemOpen
  \bibfield  {author} {\bibinfo {author} {\bibfnamefont {Y.-C.}\ \bibnamefont
  {Lin}}, \bibinfo {author} {\bibfnamefont {H.-P.}\ \bibnamefont {Komsa}},
  \bibinfo {author} {\bibfnamefont {C.-H.}\ \bibnamefont {Yeh}}, \bibinfo
  {author} {\bibfnamefont {T.}~\bibnamefont {Bj\"{o}rkman}}, \bibinfo {author}
  {\bibfnamefont {Z.-Y.}\ \bibnamefont {Liang}}, \bibinfo {author}
  {\bibfnamefont {C.-H.}\ \bibnamefont {Ho}}, \bibinfo {author} {\bibfnamefont
  {Y.-S.}\ \bibnamefont {Huang}}, \bibinfo {author} {\bibfnamefont {P.-W.}\
  \bibnamefont {Chiu}}, \bibinfo {author} {\bibfnamefont {A.~V.}\ \bibnamefont
  {Krasheninnikov}},\ and\ \bibinfo {author} {\bibfnamefont {K.}~\bibnamefont
  {Suenaga}},\ }\bibfield  {title} {\bibinfo {title} {{Single-layer ReS$_2$:
  two-dimensional semiconductor with tunable in-plane anisotropy}},\ }\href
  {https://doi.org/10.1021/acsnano.5b04851} {\bibfield  {journal} {\bibinfo
  {journal} {ACS Nano}\ }\textbf {\bibinfo {volume} {9}},\ \bibinfo {pages}
  {11249} (\bibinfo {year} {2015})}\BibitemShut {NoStop}%
\bibitem [{\citenamefont {Aslan}\ \emph {et~al.}(2015)\citenamefont {Aslan},
  \citenamefont {Chenet}, \citenamefont {van~der Zande}, \citenamefont {Hone},\
  and\ \citenamefont {Heinz}}]{Aslan2015}%
  \BibitemOpen
  \bibfield  {author} {\bibinfo {author} {\bibfnamefont {B.}~\bibnamefont
  {Aslan}}, \bibinfo {author} {\bibfnamefont {D.~A.}\ \bibnamefont {Chenet}},
  \bibinfo {author} {\bibfnamefont {A.~M.}\ \bibnamefont {van~der Zande}},
  \bibinfo {author} {\bibfnamefont {J.~C.}\ \bibnamefont {Hone}},\ and\
  \bibinfo {author} {\bibfnamefont {T.~F.}\ \bibnamefont {Heinz}},\ }\bibfield
  {title} {\bibinfo {title} {{Linearly polarized excitons in single-and
  few-layer ReS$_2$ crystals}},\ }\href
  {https://doi.org/10.1021/acsphotonics.5b00486} {\bibfield  {journal}
  {\bibinfo  {journal} {ACS Photonics}\ }\textbf {\bibinfo {volume} {3}},\
  \bibinfo {pages} {96} (\bibinfo {year} {2015})}\BibitemShut {NoStop}%
\bibitem [{\citenamefont {Qiao}\ \emph {et~al.}(2016)\citenamefont {Qiao},
  \citenamefont {Wu}, \citenamefont {Zhou}, \citenamefont {Qiao}, \citenamefont
  {Shi}, \citenamefont {Chen}, \citenamefont {Zhang}, \citenamefont {Zhang},
  \citenamefont {Ji},\ and\ \citenamefont {Tan}}]{Qiao2016}%
  \BibitemOpen
  \bibfield  {author} {\bibinfo {author} {\bibfnamefont {X.-F.}\ \bibnamefont
  {Qiao}}, \bibinfo {author} {\bibfnamefont {J.-B.}\ \bibnamefont {Wu}},
  \bibinfo {author} {\bibfnamefont {L.}~\bibnamefont {Zhou}}, \bibinfo {author}
  {\bibfnamefont {J.}~\bibnamefont {Qiao}}, \bibinfo {author} {\bibfnamefont
  {W.}~\bibnamefont {Shi}}, \bibinfo {author} {\bibfnamefont {T.}~\bibnamefont
  {Chen}}, \bibinfo {author} {\bibfnamefont {X.}~\bibnamefont {Zhang}},
  \bibinfo {author} {\bibfnamefont {J.}~\bibnamefont {Zhang}}, \bibinfo
  {author} {\bibfnamefont {W.}~\bibnamefont {Ji}},\ and\ \bibinfo {author}
  {\bibfnamefont {P.-H.}\ \bibnamefont {Tan}},\ }\bibfield  {title} {\bibinfo
  {title} {{Polytypism and unexpected strong interlayer coupling in
  two-dimensional layered ReS$_2$}},\ }\href
  {https://doi.org/10.1039/c6nr01569g} {\bibfield  {journal} {\bibinfo
  {journal} {Nanoscale}\ }\textbf {\bibinfo {volume} {8}},\ \bibinfo {pages}
  {8324} (\bibinfo {year} {2016})}\BibitemShut {NoStop}%
\bibitem [{\citenamefont {Wei}\ and\ \citenamefont {Zhang}(2002)}]{Wei2002}%
  \BibitemOpen
  \bibfield  {author} {\bibinfo {author} {\bibfnamefont {S.-H.}\ \bibnamefont
  {Wei}}\ and\ \bibinfo {author} {\bibfnamefont {S.~B.}\ \bibnamefont
  {Zhang}},\ }\bibfield  {title} {\bibinfo {title} {{Chemical trends of defect
  formation and doping limit in II-VI semiconductors: The case of CdTe}},\
  }\href {https://doi.org/10.1103/physrevb.66.155211} {\bibfield  {journal}
  {\bibinfo  {journal} {Phys. Rev. B}\ }\textbf {\bibinfo {volume} {66}},\
  \bibinfo {pages} {155211} (\bibinfo {year} {2002})}\BibitemShut {NoStop}%
\bibitem [{\citenamefont {Dey}\ \emph {et~al.}(2022)\citenamefont {Dey},
  \citenamefont {Chowdhury}, \citenamefont {Kumar},\ and\ \citenamefont
  {Kumar~Singh}}]{Dey2022}%
  \BibitemOpen
  \bibfield  {author} {\bibinfo {author} {\bibfnamefont {M.}~\bibnamefont
  {Dey}}, \bibinfo {author} {\bibfnamefont {S.}~\bibnamefont {Chowdhury}},
  \bibinfo {author} {\bibfnamefont {S.}~\bibnamefont {Kumar}},\ and\ \bibinfo
  {author} {\bibfnamefont {A.}~\bibnamefont {Kumar~Singh}},\ }\bibfield
  {title} {\bibinfo {title} {{Quantum confinement effect on defect level of
  hydrogen doped rutile VO$_2$ nanowires}},\ }\href
  {https://doi.org/10.1063/5.0095834} {\bibfield  {journal} {\bibinfo
  {journal} {J. Appl. Phys.}\ }\textbf {\bibinfo {volume} {131}},\ \bibinfo
  {pages} {235702} (\bibinfo {year} {2022})}\BibitemShut {NoStop}%
\bibitem [{\citenamefont {Zhu}\ \emph {et~al.}(2020)\citenamefont {Zhu},
  \citenamefont {Yang},\ and\ \citenamefont {Gong}}]{Zhu2020}%
  \BibitemOpen
  \bibfield  {author} {\bibinfo {author} {\bibfnamefont {G.-J.}\ \bibnamefont
  {Zhu}}, \bibinfo {author} {\bibfnamefont {J.-H.}\ \bibnamefont {Yang}},\ and\
  \bibinfo {author} {\bibfnamefont {X.-G.}\ \bibnamefont {Gong}},\ }\bibfield
  {title} {\bibinfo {title} {{Self-consistently determining structures of
  charged defects and defect ionization energies in low-dimensional
  semiconductors}},\ }\href {https://doi.org/10.1103/physrevb.102.035202}
  {\bibfield  {journal} {\bibinfo  {journal} {Phys. Rev. B}\ }\textbf {\bibinfo
  {volume} {102}},\ \bibinfo {pages} {035202} (\bibinfo {year}
  {2020})}\BibitemShut {NoStop}%
\bibitem [{\citenamefont {Arora}\ \emph {et~al.}(2017)\citenamefont {Arora},
  \citenamefont {Noky}, \citenamefont {Dr\"{u}ppel}, \citenamefont {Jariwala},
  \citenamefont {Deilmann}, \citenamefont {Schneider}, \citenamefont {Schmidt},
  \citenamefont {Del Pozo-Zamudio}, \citenamefont {Stiehm}, \citenamefont
  {Bhattacharya}, \citenamefont {Kr\"{u}ger}, \citenamefont {Michaelis~de
  Vasconcellos}, \citenamefont {Rohlfing},\ and\ \citenamefont
  {Bratschitsch}}]{Arora2017}%
  \BibitemOpen
  \bibfield  {author} {\bibinfo {author} {\bibfnamefont {A.}~\bibnamefont
  {Arora}}, \bibinfo {author} {\bibfnamefont {J.}~\bibnamefont {Noky}},
  \bibinfo {author} {\bibfnamefont {M.}~\bibnamefont {Dr\"{u}ppel}}, \bibinfo
  {author} {\bibfnamefont {B.}~\bibnamefont {Jariwala}}, \bibinfo {author}
  {\bibfnamefont {T.}~\bibnamefont {Deilmann}}, \bibinfo {author}
  {\bibfnamefont {R.}~\bibnamefont {Schneider}}, \bibinfo {author}
  {\bibfnamefont {R.}~\bibnamefont {Schmidt}}, \bibinfo {author} {\bibfnamefont
  {O.}~\bibnamefont {Del Pozo-Zamudio}}, \bibinfo {author} {\bibfnamefont
  {T.}~\bibnamefont {Stiehm}}, \bibinfo {author} {\bibfnamefont
  {A.}~\bibnamefont {Bhattacharya}}, \bibinfo {author} {\bibfnamefont
  {P.}~\bibnamefont {Kr\"{u}ger}}, \bibinfo {author} {\bibfnamefont
  {S.}~\bibnamefont {Michaelis~de Vasconcellos}}, \bibinfo {author}
  {\bibfnamefont {M.}~\bibnamefont {Rohlfing}},\ and\ \bibinfo {author}
  {\bibfnamefont {R.}~\bibnamefont {Bratschitsch}},\ }\bibfield  {title}
  {\bibinfo {title} {{Highly Anisotropic in-Plane Excitons in Atomically Thin
  and Bulklike 1T'-ReSe$_2$}},\ }\href
  {https://doi.org/10.1021/acs.nanolett.7b00765} {\bibfield  {journal}
  {\bibinfo  {journal} {Nano Lett.}\ }\textbf {\bibinfo {volume} {17}},\
  \bibinfo {pages} {3202} (\bibinfo {year} {2017})}\BibitemShut {NoStop}%
\end{thebibliography}

\begin{mcitethebibliography}{3}
\providecommand*\natexlab[1]{#1}
\providecommand*\mciteSetBstSublistMode[1]{}
\providecommand*\mciteSetBstMaxWidthForm[2]{}
\providecommand*\mciteBstWouldAddEndPuncttrue
  {\def\EndOfBibitem{\unskip.}}
\providecommand*\mciteBstWouldAddEndPunctfalse
  {\let\EndOfBibitem\relax}
\providecommand*\mciteSetBstMidEndSepPunct[3]{}
\providecommand*\mciteSetBstSublistLabelBeginEnd[3]{}
\providecommand*\EndOfBibitem{}
\mciteSetBstSublistMode{f}
\mciteSetBstMaxWidthForm{subitem}{(\alph{mcitesubitemcount})}
\mciteSetBstSublistLabelBeginEnd
  {\mcitemaxwidthsubitemform\space}
  {\relax}
  {\relax}

\bibitem[Dileep \latin{et~al.}(2016)Dileep, Sahu, Sarkar, Peter, and
  Datta]{Dileep2016}
Dileep,~K.; Sahu,~R.; Sarkar,~S.; Peter,~S.~C.; Datta,~R. {Layer specific
  optical band gap measurement at nanoscale in MoS$_2$ and ReS$_2$ van der
  Waals compounds by high resolution electron energy loss spectroscopy}.
  \emph{J. Appl. Phys.} \textbf{2016}, \emph{119}, 114309\relax
\mciteBstWouldAddEndPuncttrue
\mciteSetBstMidEndSepPunct{\mcitedefaultmidpunct}
{\mcitedefaultendpunct}{\mcitedefaultseppunct}\relax
\EndOfBibitem
\bibitem[Tongay \latin{et~al.}(2014)Tongay, Sahin, Ko, Luce, Fan, Liu, Zhou,
  Huang, Ho, Yan, Ogletree, Aloni, Ji, Li, Li, Peeters, and
  Wu]{Tongay2014natcomm}
Tongay,~S. \latin{et~al.}  {Monolayer behaviour in bulk ReS$_2$ due to
  electronic and vibrational decoupling}. \emph{Nat. Commun.} \textbf{2014},
  \emph{5}, 1--6\relax
\mciteBstWouldAddEndPuncttrue
\mciteSetBstMidEndSepPunct{\mcitedefaultmidpunct}
{\mcitedefaultendpunct}{\mcitedefaultseppunct}\relax
\EndOfBibitem
\bibitem[Jariwala \latin{et~al.}(2016)Jariwala, Voiry, Jindal, Chalke, Bapat,
  Thamizhavel, Chhowalla, Deshmukh, and Bhattacharya]{Jariwala2016}
Jariwala,~B.; Voiry,~D.; Jindal,~A.; Chalke,~B.~A.; Bapat,~R.; Thamizhavel,~A.;
  Chhowalla,~M.; Deshmukh,~M.; Bhattacharya,~A. {Synthesis and characterization
  of ReS$_2$ and ReSe$_2$ layered chalcogenide single crystals}. \emph{Chem.
  Mater.} \textbf{2016}, \emph{28}, 3352--3359\relax
\mciteBstWouldAddEndPuncttrue
\mciteSetBstMidEndSepPunct{\mcitedefaultmidpunct}
{\mcitedefaultendpunct}{\mcitedefaultseppunct}\relax
\EndOfBibitem
\end{mcitethebibliography}
\end{document}